\documentclass[aps,prb,showpacs,twocolumn,floats,superscriptaddress]{revtex4-1}
\pdfoutput=1
\usepackage[utf8]{inputenc}
\usepackage[pdftex]{graphicx}
\usepackage{amssymb}
\usepackage{amsmath}
\usepackage{color}
\usepackage{dcolumn}
\usepackage{bm}
\usepackage[caption=false]{subfig}
\usepackage{placeins}
\usepackage{subfig}
\usepackage{paralist}
\usepackage{tabularx}
\usepackage[english]{babel}

\renewcommand\vec[1]{\ensuremath\boldsymbol{#1}} 
\newcommand{\vsigma}{\vec{\sigma}}
\newcommand{\vq}{\vec{q}}

\newcommand{\vspin}{\hat{\vec{S}}}

\RequirePackage[
   hyperindex,colorlinks,bookmarksnumbered,
   plainpages=true,pdfstartview=FitH]{hyperref}
\hypersetup{linkcolor=blue,urlcolor=blue,citecolor=blue}
\usepackage{hyperref}

\definecolor{darkgreen}{rgb}{0,0.5,0}
\definecolor{darkblue}{rgb}{0,0,0.5}
\definecolor{purple}{rgb}{0.35,0,0.35}
\definecolor{orange}{rgb}{1,0.5,0}
\definecolor{wbcolor}{rgb}{0,.6,1}
\definecolor{todocolor}{rgb}{1,0,0}
\definecolor{jvdcolor}{rgb}{0,0,1}

\newcommand{\Eq}[1]{Eq.~(\ref{#1})}
\newcommand{\Eqs}[1]{Eqs.~(\ref{#1})}

\newcommand{\Sec}[1]{Sec.~\ref{#1}}

\newcommand{\Ref}[1]{Ref.~[\onlinecite{#1}]}

\newcommand{\Fig}[1]{Fig.~\ref{#1}}

\newcommand{\Figs}[1]{Figs.~\ref{#1}}

\graphicspath{ {Figures/}}

\begin{document}

\title{Symmetric minimally entangled typical thermal states}
\author{Benedikt Bruognolo}
\author{Jan von Delft}
\author{Andreas Weichselbaum}
\affiliation{Physics Department, Arnold Sommerfeld Center for Theoretical Physics, and Center for NanoScience,
Ludwig-Maximilians-Universit\"at, Theresienstra{\ss}e 37, 80333 M\"unchen, Germany}

\date{\today}

\begin{abstract}
We extend White's minimally entangled typically thermal states approach (METTS) to allow Abelian and non-Ablian symmetries to be exploited when computing finite-temperature response functions in one-dimensional (1D) quantum systems. Our approach, called SYMETTS, starts from a METTS sample of states that are not symmetry eigenstates, and generates from each a symmetry eigenstate. These symmetry states are then used to calculate dynamic response functions. SYMETTS is ideally suited to determine the low-temperature spectra of 1D quantum systems with high resolution.
We employ this method to study a generalized diamond chain model for the natural mineral azurite Cu${}_3$(CO${}_3)_2$(OH$)_2$, which features a plateau at $\frac{1}{3}$ in the magnetization curve at low temperatures. Our calculations provide new insight into the effects of temperature on magnetization and excitation spectra in the plateau phase, which can be fully understood in terms of the microscopic model.
\end{abstract}

\maketitle


\section{Introduction} \label{sec:intro}
The simulation of dynamical quantities in one-dimensional (1D) quantum many-body systems still poses a major challenge for theoretical condensed matter physics, particularly at finite temperature. From an experimentalist's perspective, there is high demand for such calculations for a variety of reasons: (i) Experimental measurements hardly allow to study solely ground-state physics as thermal fluctuations cannot be eliminated altogether. Thus for a direct comparison with experimental data, it is essential to include temperature in the theoretical modeling. (ii) Technical advances have nowadays drastically enhanced the precision of neutron scattering and electron resonance spectroscopy, which for example allows the measurement of dynamic observables such as momentum-resolved excitation spectra in effective 1D materials with very high resolution.\cite{Zaliznyak_PRL_2003,zaliznyak2005magnetic,lake2005quantum,Rule_PRL_2008,rule2011dynamics,Zvyagin_LTP_2012}  (iii) Thermal fluctuations can cause new phenomena, which are not captured by the ground-state physics of the system. Two examples are the sudden emergence of a single spinon dispersion (``Villain mode'') in XXZ-like spin-chain materials,\cite{Villain19751,Nagler_PRB_83,James_PRB_2009} or the existence of quantum critical phases in various strongly correlated materials.\cite{sachdev2001quantum} 

Which numerical tools can be employed to simulate such dynamic observables in a 1D quantum system? At zero temperature, the density matrix renormalization group (DMRG) is the most successful exact numerical method for describing quantum many-body systems regarding their static and dynamic ground-state properties.\cite{White92,Schollwoeck2011} DMRG based algorithms have also been successfully extended to treat systems at finite temperature, yet the computational efficiency of such approaches is still limited. Exact diagonalization (ED) or quantum Monte Carlo (QMC)\cite{Suzuki_QMC_1977,Hirsch_PRB_1982,Sandvik_PRB_1991} can only be considered as complementary approaches rather than proper alternatives to DMRG, since the applicability of ED is restricted by small system sizes, and that of QMC by the need for performing an ill-defined analytic continuation, and often also by the occurrence of a sign problem.  Thus, the simulation of experimentally relevant quantities such as dynamic response functions represents a highly demanding and difficult task for finite-temperature numerics.

Whereas early DMRG approaches for computing finite-temperature response functions  for a 1D quantum system have been based on the transfer matrix renormalization group (TMRG),\cite{Nishino_JPSJ_1995,Bursill_JPCM_1996,Shibata_JPSC_1997,Wang_PRB_1997} today the most popular method builds on the purification of the density matrix  in the matrix-product-state (MPS) formalism.\cite{Verstraete_PRL_2004} The response functions can then be calculated with high precision by using tDMRG in the real-time realm, and a subsequent Fourier transform also allows the computation of spectral functions.\cite{WhiteFeiguin04,White_PRB_2008_LinPre,Barthel_PRB_2009,Karrasch_PRL_2012,barthel_2012scaling,lake2013multispinon,Barthel_NJP_2013} Purification can also be combined with a Chebyshev expansion technique to determine finite-temperature spectral functions directly in frequency space.\cite{Tiegel_PRB_2014, Holzner_PRB_2011} Although these methods have been successfully applied to a number of experimental setups, the accessible time scale (or maximal Chebyshev expansion order) and hence the spectral resolution is limited, as the propagation of excitations during the dynamic evolution yields a linear growth of entanglement, leading to an exponential increase in the required numerical resources. In addition, the encoding of mixed states inevitably requires doubling the size of the Hilbert space and introduces additional entanglement between the physical state and its environment, which limits the efficiency of purification simulations towards low temperatures.

An alternative way to compute finite-temperature quantities was recently presented by White in \Ref{White_PRL_2011}. Instead of purifying the density matrix, an ensemble of pure states is introduced that are constructed to resemble the typical state of a quantum system at finite temperature. It has been shown that these so-called minimally entangled typical thermal states (METTS) excellently represent the thermal properties of the system of interest. At the same time, they can efficiently be  represented in the MPS formalism as their entanglement is very low. \cite{White_PRL_2011,Stoudenmire_NJP_2010} The METTS approach was originally only used to compute static quantities of spin-chains\cite{White_PRL_2011,Stoudenmire_NJP_2010} and fermions. \cite{Alvarez_PRB_2013} In the meantime, it has also been applied to simulate finite-temperature quenches\cite{Bonnes_PRL_2014} and response functions.\cite{binder2014minimally}

The numerical effort for constructing a single METTS is comparable to ground-state DMRG, since METTS avoids the explicit computation of the density matrix. Since METTS calculations are also easily parallelized, it has  originally been considered to be a more efficient finite-temperature formulation than purification. More recently, \Ref{binder2014minimally} showed in a detailed study that this claim cannot generally be supported, because the additional statistical error source introduced by the sampling
  increases computational costs, especially at high temperatures. Nevertheless, METTS still offers much potential towards the simulations of low-temperature properties of complex models, as long as one does not insist on reducing the statistical error to be as small as the truncation error.

To bring out the full potential of METTS for the calculation of dynamic quantities, this work addresses a severe constraint of the current formulation of the algorithm: the ensemble states cannot be chosen such that they respect inherent \emph{symmetries} of the system and at the same time minimize autocorrelation effects. This drastically increases the numerical resources necessary for computing the real-time evolution of the ensemble states, as the MPS have not been decomposed into symmetry blocks by means of the symmetry-induced selection rules. \cite{McCulloch_EPL_2002,weichselbaum12a,Singh_PRB_2012,Singh_PRB_2013} To remedy this problem, we introduce an intuitive and easily implementable extension of White's approach: starting from a METTS sample of states that are not symmetry eigenstates, we generate a sample of symmetry eigenstates, called SYMETTS. These states allow both simple Abelian and more complex non-Abelian symmetries to be exploited in the computation of dynamic quantities.

As an experimentally relevant application of SYMETTS, we study temperature effects on the $\frac{1}{3}$ magnetization plateau of a generalized diamond chain model, which has been derived as a microscopic model for the natural mineral azurite $\text{Cu}_3(\text{CO}_3)_2(\text{OH})_2$.\cite{jeschke2011multistep,Honecker_JCM_2011} This material has attracted much attention due to the discovery of a plateau at $\frac{1}{3}$ in the magnetization curve at low temperatures.\cite{kikuchi2003magnetic,Kikuchi_PRL_2005,GU_PRLCom_2006,Gu_PRB_2007,Mikeska_PRB_2008,Rule_PRL_2008,Kang_JCMP_2009,jeschke2011multistep,Honecker_JCM_2011,rule2011dynamics,Ananikian_EPJB_2012,Cong_PRB_2014,Richter201539} Via real-time evolution of SYMETTS ensembles, it is possible to obtain highly resolved excitation spectra in the $\frac{1}{3}$ plateau phase for various temperatures. We observe a crossing of monomer and dimer branches with increasing magnetic field, which intuitively explains the effects of finite temperature on the magnetization in the plateau phase.

The paper is organized as follows. In \Sec{sec:METTS} we briefly review the original METTS algorithm and the necessity of choosing a symmetry-breaking collapse routine to generate the ensemble. \Sec{sec:SYMETTS} introduces a METTS formulation based on symmetry eigenstates for models with both Abelian and non-Abelian symmetries. \Sec{sec:results1} summarizes benchmark calculations for static and dynamic observables for the  spin-$\frac{1}{2}$ XXZ chain. In \Sec{sec:azurite} SYMETTS is employed to study an experimentally relevant microscopic model for the natural mineral azurite. A technical discussion on the combination of SYMETTS with a Chebyshev expansion to directly calculate dynamic correlators in frequency space is relegated to appendix \ref{sec:CheMETTS}. The computational efficiency of SYMETTS in the context azurite is assessed in appendix \ref{app:cputime}.


\section{Minimally Entangled Typical Thermal States} \label{sec:METTS}

\subsection{METTS calculations thermal quantities}
First of all, we review the construction of a METTS sample to approximate a thermal expectation value $\langle \hat{A} \rangle_{\beta}$ for a general chain model with $N$ sites. To this end, the trace of a thermal expectation value $\langle \hat{A} \rangle_{\beta} = \text{Tr}[\rho_{\beta} \hat{A}]$ is expanded in terms of an orthonormal basis $\{ |\vsigma \rangle \}$ of classical product states (CPS) of the form $| \vsigma \rangle = |\sigma^1\rangle|\sigma^2\rangle...|\sigma^N\rangle$. Each such state has an entanglement entropy of exactly zero. Thus these states  represent a natural choice for a basis at infinite temperature, where the system should behave classically. In addition, their entanglement growth under imaginary-time evolution remains comparatively low, hence the designation ``minimally entangled'' states. The expectation value of $\hat{A}$ can be written as
\begin{eqnarray}
\langle \hat{A} \rangle_{\beta} &=& \frac{1}{Z_{\beta}} \sum_{\vsigma} \langle \vsigma | e^{-\beta \hat{H}/2} \hat{A}  e^{-\beta \hat{H}/2} |\vsigma \rangle \nonumber\\ 
 &=&  \frac{1}{Z_{\beta}} \sum_{\vsigma} P_{\vsigma} \langle \phi_{\vsigma} |  \hat{A} | \phi_{\vsigma} \rangle, \label{eq:expT}
\end{eqnarray}
with the partition function $  Z_{\beta} = \text{Tr}[e^{-\beta \hat{H}}] =   \sum_{\vsigma} P_{\vsigma}  $. The normalized states $|\phi_{\vsigma}  \rangle$ represent a set of METTS with corresponding probabilities $P_{\vsigma}$, defined as
\begin{equation}
|\phi_{\vsigma}  \rangle = \frac{1}{\sqrt{P_{\vsigma} }} e^{-\beta \hat{H}/2}|\vsigma \rangle, \quad P_{\vsigma}  = \langle \vsigma | e^{-\beta \hat{H}} | \vsigma \rangle. \label{eq:defMETTS}
\end{equation}
By sampling the METTS $|\phi_{\vsigma}\rangle$ according to the probability distribution $P_{\vsigma}/Z_{\beta}$, the calculation of a thermal expectation value can be reformulated into taking the plain average of $\langle \phi_{\vsigma} | \hat{A} | \phi_{\vsigma} \rangle$.

To obtain a METTS sample $\{|\phi_{\vsigma}\rangle\}$  with the correct probability distribution,  a Markov chain of CPS $|\vsigma\rangle$ is generated. This is done in a way that obeys detailed balance, which guarantees reproducing the probability distribution $P_{\vsigma}/Z_{\beta}$. The sampling algorithm can be set up sequentially. To this end, one starts from an arbitrary CPS $|\vsigma\rangle$ and conducts what is called a \emph{thermal step}: \\

\begin{compactenum}[(i)]
\item A single METTS $|\phi_{\vsigma}\rangle$ is generated by evolving the CPS in imaginary time and normalizing it.
\item  A measurement of all local degrees of freedom is performed by projecting (or collapsing) $| \phi_{\vsigma} \rangle$ into  a new CPS $|\vsigma'\rangle$  with probability $p_{\vsigma' \vsigma} = | \langle \vsigma' | \phi_{\vsigma} \rangle|^2$. The transition probabilities obey detailed balance $p_{\vsigma' \vsigma} P_{\vsigma} = p_{\vsigma \vsigma'} P_{\vsigma'}$ by construction. \\ \end{compactenum}

\begin{figure}[t]
         \centering
\includegraphics[width=\linewidth]{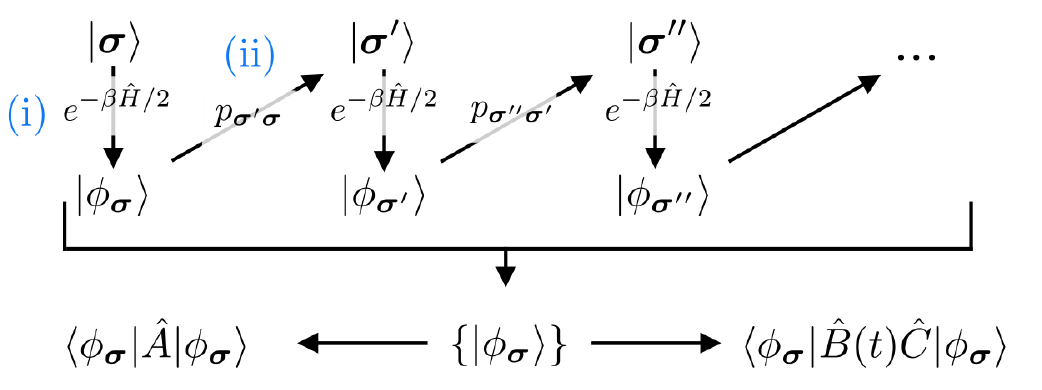}
\caption{Schematic illustration of the METTS algorithm. For details on how to explicitly evaluate response functions of the type $\langle \hat{B}(t) \hat{C} \rangle_{\beta}$ see \Sec{sec:dyn}.}
\label{fig:METTS_alg}
\end{figure}

The thermal step is then repeated with the newly generated CPS to generate a METTS (see \Fig{fig:METTS_alg} for illustration). By construction, the correct distribution is recovered as a fixed point of this procedure. To eliminate any artificial bias caused by the choice of the initial random CPS, the first few thermal steps are neglected in the calculation of any static observable  $\langle \phi_{\vsigma} |  \hat{A} | \phi_{\vsigma} \rangle$ or dynamic response function $\langle \hat{B}(t) \hat{C} \rangle_{\beta}$.

By making good choices for the local measurements (see Sec.~\ref{sec:ergo}), the sample size $M$ can be chosen surprisingly small to obtain accurate results [$M \sim \mathcal{O}(10^2$--$10^3)$]. 

At this point, a comment on accuracy is in order. It was recently pointed out by \Ref{binder2014minimally} that the purification approach drastically outperforms METTS, because for fixed total computation time it reaches more accurate results, where the accuracy was judged by comparing to quasi-exact calculations. However, this should not be a surprise, since the METTS sampling introduces an additional statistical error source, which generally scales as $ \beta^{-1}/\sqrt{M-1}$. Obviously, this prevents a perfect convergence of METTS results towards exact data and limits the efficiency at very high temperatures in comparison to a non-statistical method. Nevertheless, we believe that METTS offers much potential towards the simulations of low-temperature properties of complex models, as long as one does not insist on pushing the statistical error towards the order of the truncation error.

\subsection{Ergodicity and efficient sampling} \label{sec:ergo}
Generating a new CPS $|\vsigma\rangle$ by collapsing a METTS represents the most crucial step of the sampling algorithms, as a bad choice of measurement basis leads to a drastically increased autocorrelation time. \cite{Stoudenmire_NJP_2010}

 Let us assume that the local Hilbert space of each site $j$ in our chain model is represented by an orthonormal basis $|\sigma_j \rangle$ of size $d$, $\sigma_j \in \{1,2,...,d\}$. The projective measurement $| \phi_{\vsigma} \rangle \to |\vsigma'\rangle$ can be efficiently carried out site by site by making use of the well-defined orthogonality relations for a MPS, typically starting at one end of the chain (in our case site 1). To this end, the $d$ transition probabilities $p(\sigma_1) = \langle \phi_{\vsigma}| \hat{P}(\sigma_1) | \phi_{\vsigma}\rangle$ are calculated by introducing the projectors $\hat{P}(\sigma_1) = |\sigma_1\rangle \langle \sigma_1 | $. Then one of the $d$ states is chosen with probability $p(\sigma_1)$ by rolling a dice. The state is collapsed by the application of the projector $\hat{P}(\sigma_1)$, and the orthonormal center of the MPS is shifted to the next site, where the collapse process is repeated.
 
In principle, the orthonormal basis $|\sigma_j \rangle$ on each site $j$ can be chosen arbitrarily. Nevertheless, there are good and bad choices with respect to the sampling efficiency. We illustrate this for the example of  the  spin-$\frac{1}{2}$ XXZ Heisenberg chain, 
\begin{equation}\label{eq:XXX_H}
\hat{H} = J \sum^N_j \big[ \hat{S}_{j}^x  \hat{S}_{j+1}^x + \hat{S}_{j}^y  \hat{S}_{j+1}^y + \Delta \hat{S}_{j}^z  \hat{S}_{j+1}^z\big ] + h \sum_j^N \hat{S}_{j}^z \ , 
\end{equation}
for the isotropic case $J=1$, $\Delta=1$ and $h=0$.~This model features a non-Abelian SU$(2)_{\rm spin}$ symmetry, which can be reduced to an Abelian U(1) symmetry, e.g., by considering the total magnetization $S_{\rm tot}^z$ as a good quantum number. At first sight, the eigenstates of the spin operator $\hat{S}^z_j$ resemble a natural choice for the orthonormal basis set $|\sigma_j \rangle$, since this choice allows the encoding of the projectors in the form of diagonal operators. Moreover, all resulting CPS are eigenstates of $\hat{S}_{\rm tot}^z = \sum_j^N \hat{S}^z_j $. Therefore, it is possible to directly implement the Abelian U(1) symmetry in the MPS representation resulting in a massive reduction of computational effort.

However, a collapse routine based on measurements along the $z$ axis only (``$z$ collapse'') leads to a serious problem with ergodicity, as already extensively discussed in \Ref{Stoudenmire_NJP_2010}. Subsequently generated CPS are strongly correlated and thus the autocorrelation times are very long, so that the bias arising from the initial random CPS cannot be removed in a few thermal steps. Additionally, CPS generated from subsequent thermal steps always have the same total magnetization $S_{\rm tot}^z$ since the $z$ collapse conserves this quantity. It is therefore impossible to cover different $S_{\rm tot}^z$ sectors with the sampling algorithm described above, as one is always stuck in the symmetry sector of the initially chosen CPS.   

This issue can be resolved by randomly choosing a different local basis for each site of the chain (``random collapse''). Alternatively, it has been shown that alternating in subsequent thermal steps between two basis sets that are maximally mixed relative to each other, e.g., the eigenstates of $\hat{S}^z$ and $\hat{S}^x$,  also restores ergodicity and covers multiple symmetry sectors of the sample (``maximally mixed collapse'').  

\begin{figure}[h!]
\centering
\includegraphics[width=\linewidth]{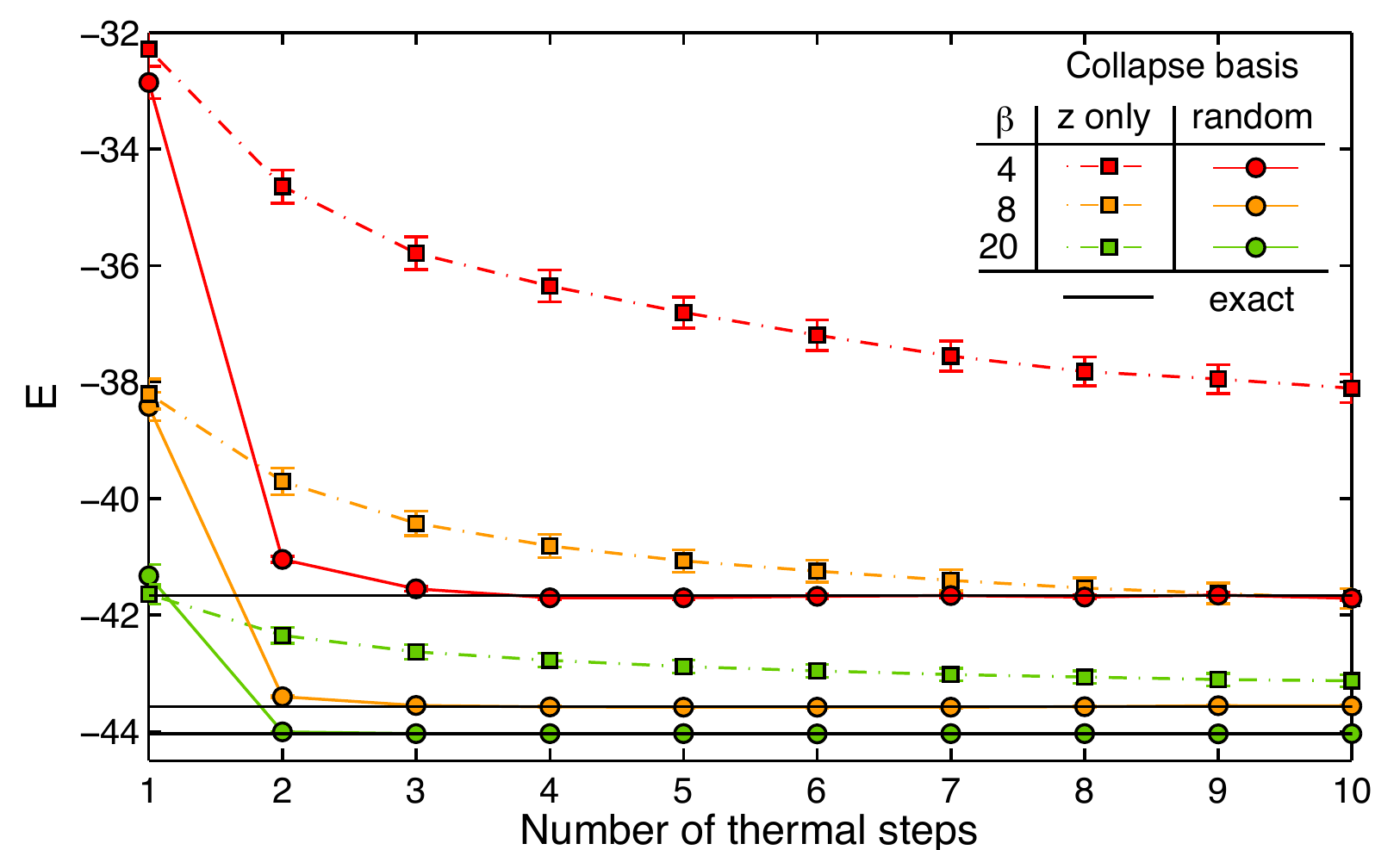}
\vspace{-10pt}
\caption{Energy of the spin-$\frac{1}{2}$ Heisenberg chain with $N=100$ spins for $\beta=4,8,20$. Details on the setup of the imaginary-time evolution can be found in \Sec{sec:results1}. Starting from an ensemble of 100 randomly generated CPS, we conduct ten thermal steps with each state and measure the ensemble average of the total energy in each step. The different basis choices for the CPS collapse become apparent in the autocorrelation times. Whereas measuring along the $z$ axis only (squares) leads to strong autocorrelations that prevent the energy to converge towards the exact value (black lines), randomly chosen measurement bases (circles) result in short autocorrelation times of a few thermal steps.\cite{Stoudenmire_NJP_2010}}
\label{fig:szvsrand}
\end{figure}

We illustrate the failure of  the $z$ collapse routine by starting from an ensemble of randomly generated CPS and then conducting 10 thermal steps with each state of the ensemble for three different values of $\beta$. After each step, we measure the ensemble average of the total energy, which is displayed in \Fig{fig:szvsrand}. For comparison, we also calculate the same quantity using the random collapse routine. When choosing a random basis for each CPS collapse (circles), the total energy of the ensemble is already well converged towards the exact value after a few thermal steps, because autocorrelations between subsequent CPS are practically absent.

In contrast, when measuring along the $z$ axis only (squares), the total energy is nowhere near its exact value, even after 10 thermal steps. We have discussed the causes of this  behavior above: first of all, one can identify strong correlations between subsequent CPS during the application of the $z$ collapse resulting in an increase of autocorrelation time. In addition, each CPS remains in its initial  symmetry sector. If the different symmetry sectors are not distributed according to the correct probability distribution at a specific value $\beta$ (which is very unlikely starting from a random set), the ensemble cannot capture the correct behavior of the system, as the sample is biased towards specific sectors. This explains why the average energy is not only converging slowly towards the exact value, but rather seems to saturate at a significantly higher value. Thus, a symmetry conserving collapse routine that is based on measurements in a fixed local basis is clearly impracticable.

If we want to retain the ergodicity of the METTS sample, we are left to choose between the random or the maximally mixed collapse routine. This comes at a price, as the ensemble states cannot be chosen such that they conserve inherent symmetries of the system, because both collapse routines clearly require symmetry-breaking measurements. However, the efficient treatment of symmetries is often essential for calculating especially dynamic properties of complex models, such as 1D systems and 2D lattice models with experimental relevance. Since the current METTS setup does not allow Abelian or non-Abelian symmetries to be exploited, it is not suitable for accessing dynamic observables of such complex systems.

In the following, we show how to resolve this fundamental issue by a simple extension of the sampling algorithm that will enable us to systematically build a METTS ensemble based on symmetry eigenstates.


\section{Symmetric METTS} \label{sec:SYMETTS}
\subsection{Symmetries}
\label{sec:sym}
The matrix-product-state framework allows for a straightforward incorporation of  symmetries of the model Hamiltonian.\cite{McCulloch_EPL_2002,weichselbaum12a,Singh_PRB_2012,Singh_PRB_2013} Generally speaking, the symmetry-induced selection rules cause a large number of matrix elements to be exactly zero, thus bringing the Hamiltonian into a block-diagonal structure and subdividing tensors into well-defined symmetry sectors. Keeping only the non-zero elements, we can achieve tremendous improvement in speed and accuracy in numerical simulations by the inclusion of symmetries. In the context of non-Abelian symmetries, the non-zero data blocks are not independent of each other and can be further compressed using the Clebsch-Gordan algebra for multiplet spaces. Here we refrain from discussing this topic at length and refer to \Ref{weichselbaum12a} for a detailed review on the treatment of symmetries in tensor network applications.

Following the notation of \Ref{weichselbaum12a}, we label the state space in terms of the symmetry eigenbasis $|qn;q_z\rangle$, where the quantum labels $q$ denotes the irreducible representation of the symmetry group $\mathcal{S}$ of the Hamiltonian $\hat{H}$. Every symmetry generator $\hat{S}_{\alpha}$ satisfies $[\hat{H},\hat{S}_{\alpha}]=0$ . Hence, all states in a given Hilbert space corresponding to a certain $q$-label are combined into a symmetry block $q$. The label $n$ identifies a particular multiplet within the specific symmetry block $q$. The internal multiplet label $q_z$ resolves the internal structure of the corresponding multiplet. In the context of Abelian symmetries  the Clebsch-Gordan structure becomes trivial, hence the $q_z$ labels take the role of $q$ labels. Note that this notation can be easily generalized to the treatment of multiple symmetries.\cite{weichselbaum12a}

To further clarify the notation, we consider the example of the isotropic Heisenberg chain in \Eq{eq:XXX_H}, which features a SU$(2)_{\rm spin}$ symmetry, $\mathcal{S} = \text{SU}(2)_{\rm spin}$. We make the usual choice of basis in which the $z$ component of the spin operator, $\hat{S}_z$ is diagonal and label a general spin multiplet by $|q , q_z \rangle \equiv | S , S_z \rangle $. The spin multiplet label can take the values $q = 0, \frac{1}{2}, 1 , \frac{3}{2}, ...$, while the internal multiplet label, corresponding to the $z$ component of the spin, is restricted to $q_z \in \{-q , -q+1 , ... , + q \}$. 

Now consider a typical MPS scenario, where the wave function $|\psi\rangle$ in the local picture of site $j$ can be represented as 
\begin{equation} \label{eq:mps}
|\psi \rangle = \sum_{L\sigma_j R} A^{[\sigma_j]}_{LR} |L\rangle |\sigma_j\rangle |R\rangle .
\end{equation}
In the presence of symmetries, the physical state space at site $j$ as well as the left and right orthonormal basis states can be written as  $|L\rangle \equiv |ql;q_z\rangle$, $| \sigma_j\rangle \equiv |q'm; q'_z\rangle$, $|R\rangle \equiv |q''n;q''_z\rangle$. Hence, symmetry labels can be introduced naturally in the MPS representation. In particular, every leg or bond in the usual diagrammatic depiction of a MPS can be assigned a multiplet label, here $q, q'$ and $q''$, e.g.,
\begin{equation}
A^{[q']}_{qq''}  =  \parbox[h][0.1\linewidth][c]{0.20\linewidth}{
        \includegraphics[width=\linewidth]{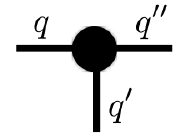}} , 
\end{equation}

\subsection{METTS with symmetry eigenstates}\label{sec:SPS}
In order to work with a symmetry conserving METTS ensemble, we reformulate \Eqs{eq:expT} and (\ref{eq:defMETTS}) in terms of symmetry eigenstates before introducing an efficient sampling routine (see Sec.~\ref{sec:sampling}). In place of the CPS we introduce a set of symmetry product states (SPS) $|\vq \rangle$, that can be considered as symmetrized counterparts of the CPS. A SPS is a MPS with (multiplet) bond dimension one, where each bond represents a single, unique symmetry block $q_j$, and that block contains just a single multiplet ($n_j=1$). Thus the SPS can be fully characterized by a set of $N$ quantum labels $\vq = \{q_1, q_2, q_3, ..., q_N\}$, one label $q_j$ per site/bond $j$ labeling the corresponding symmetry sector. The overall symmetry sector of each SPS $|\vq \rangle$ is fully determined by the $q$-label of the last bond, $q_N$. 

\begin{figure}[t]
         \centering
\includegraphics[width=\linewidth]{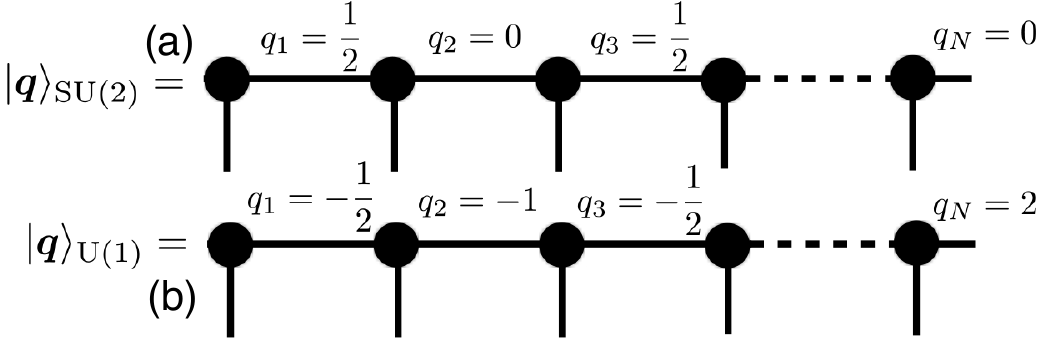}
\caption{Schematic illustration of the SPS (a) in \Eq{eq:SPS_SU2} and (b) in \Eq{eq:SPS_U1}.}
\label{fig:SPS}
\end{figure}

The simplest example of a SPS for the SU(2) symmetric Heisenberg chain is to combine pairs of neighboring spins to singlets
\begin{eqnarray} \label{eq:SPS_SU2}
| \vq \rangle_{\rm SU(2)} &=& (|\uparrow_1\rangle |\downarrow_2 \rangle - |\uparrow_2\rangle |\downarrow_1 \rangle) (|\uparrow_3\rangle |\downarrow_4 \rangle - |\uparrow_3\rangle |\downarrow_4 \rangle)  \nonumber \\
&& ... \ (|\uparrow_{N-1}\rangle |\downarrow_N \rangle - |\uparrow_{N-1}\rangle |\downarrow_N \rangle). 
\end{eqnarray}
For spin-$\frac{1}{2}$ systems, the quantum labels $\vq$ correspond to a sequence of $q_j=\frac{1}{2},0$ for odd  and  even bonds, respectively, with a total spin $q_N \equiv S_{\rm tot} = 0$, as illustrated in \Fig{fig:SPS}(a). Although the (multiplet) dimension on each bond remains one for a non-Abelian SPS, it is no longer a pure product state in the classical sense as the internal multiplet structures can introduce non-trivial entanglement between neighboring sites. 

The same formalism also applies for Abelian SPS.  Reducing SU(2) to an Abelian U(1) symmetry in the Heisenberg model, e.g., by choosing an anisotropy $\Delta \neq 1$ in \Eq{eq:XXX_H} or adding a finite magnetic field in the $z$ direction, and choosing the total spin $S_{\rm tot}^z$ as conserved quantity, a typical SPS takes the form of 
\begin{equation} \label{eq:SPS_U1}
| \vq\rangle_{\rm U(1)} = |\downarrow_1\rangle |\downarrow_2 \rangle |\uparrow_3 \rangle |\downarrow_4 \rangle ... |\uparrow_N \rangle.
\end{equation}
In this case, the quantum label $q_j$ represents the sum of all $S^z$-contributions for sites $i\leq j$, i.e., $q_j= \sum_{i\leqslant j} S_i^z$ as shown in \Fig{fig:SPS}(b). Hence, the total magnetization $S_{\rm tot}^z$ of the SPS is given by the last label $q_N$. In the Abelian case, a SPS can be understood as a direct product of local symmetry eigenstates of each site, and hence is always represented by a MPS of bond dimension one, much like a classical product state.

Analogously to the CPS basis set $\{ |\vsigma \rangle \}$, a full set of SPS  $\{ |\vq \rangle \}$ represents a complete orthonormal basis taking into account all possible symmetry sectors of the system. Thus, we can proceed as above and expand the trace of a thermal expectation value $\langle \hat{A} \rangle_{\beta} = \text{Tr}[\rho_{\beta} \hat{A}]$ in terms of the symmetry product states $|\vq\rangle$:
\begin{eqnarray}
\langle \hat{A} \rangle_{\beta} &=& \frac{1}{Z_{\beta}} \sum_{ \vq} \langle \vq | e^{-\beta \hat{H}/2} \hat{A}  e^{-\beta \hat{H}/2} |\vq \rangle \nonumber\\ 
 &=&  \frac{1}{Z_{\beta}}  \sum_{\vq} P_{\vq} \langle \phi_{\vq} |  \hat{A} | \phi_{\vq} \rangle. \label{eq:expT_SPS} 
\end{eqnarray}
The normalized states $| \phi_{\vq} \rangle $ now represent a set of symmetric METTS (SYMETTS) with probabilities $P_{\vq}$  defined in analogy to \Eq{eq:defMETTS},
\begin{subequations}
\label{eq:defSYMETTS}
\begin{eqnarray}
| \phi_{\vq} \rangle &=& \frac{1}{\sqrt{P_{\vq}}} e^{-\beta \hat{H}/2}|\vq \rangle, \\
 \quad P_{\vq} &=& \langle\vq| e^{-\beta \hat{H}} |\vq \rangle\ .
\end{eqnarray}
\end{subequations}

The thermal expectation value $\langle \hat{A} \rangle_{\beta}$ is now estimated by sampling SYMETTS  $| \phi_{\vq} \rangle $ according to the probability distribution $P_{\vq}/Z_{\beta}$. However, we still have to establish how to sample a set of SYMETTS $\{ | \phi_{\vq} \rangle \}$ according to the correct probability distribution $P_{\vq}/Z_{\beta}$, in a way that ensures ergodicity.

\subsection{Algorithm for efficient sampling}
\label{sec:sampling}
We illustrated in Sec.~\ref{sec:ergo} for the spin-$\frac{1}{2}$ Heisenberg chain that a collapse routine purely based on measurements along the $z$ axis, conserving the U$(1)_{\rm spin}$  symmetry, fails to capture the correct thermal properties of the model. This is due to strong autocorrelation effects and the fact that the symmetry sectors initially are distributed randomly instead of according to the correct probability distribution  $P_{\vq}/Z_{\beta}$. From this discussion, we can learn that the SYMETTS sample has to be generated from SPS that already capture the correct distribution of symmetry sectors. \footnote{It was briefly noted in \Ref{Stoudenmire_NJP_2010} that this can already be achieved using a maximally mixed collapse procedure and treating  the $x$ basis as an effective  $z$ basis. This is possible due to the presence of the SU$(2)_{\rm spin}$ symmetry in the isotropic model which is effectively reduced to U(1) by this implicit switch of bases. Note that this is not possible in absence of SU(2), e.g., in the anisotropic XXZ model or in the presence of a magnetic field. Hence this trick cannot be used to exploit the full symmetry of the respective model.} 

This can be achieved by starting from a typical thermal state, which already incorporates all the necessary thermal information. To this end, we extend  the METTS sampling algorithm. After the conduction of a thermal step with a non-symmetric CPS $|\vsigma\rangle$ and METTS $|\phi_{\vsigma} \rangle$ using a random or maximally mixed collapse [c.f~(i) and (ii) in \Sec{sec:METTS}], we employ an additional \emph{symmetrization step}:\\

\begin{figure}[t]
         \centering
\includegraphics[width=\linewidth]{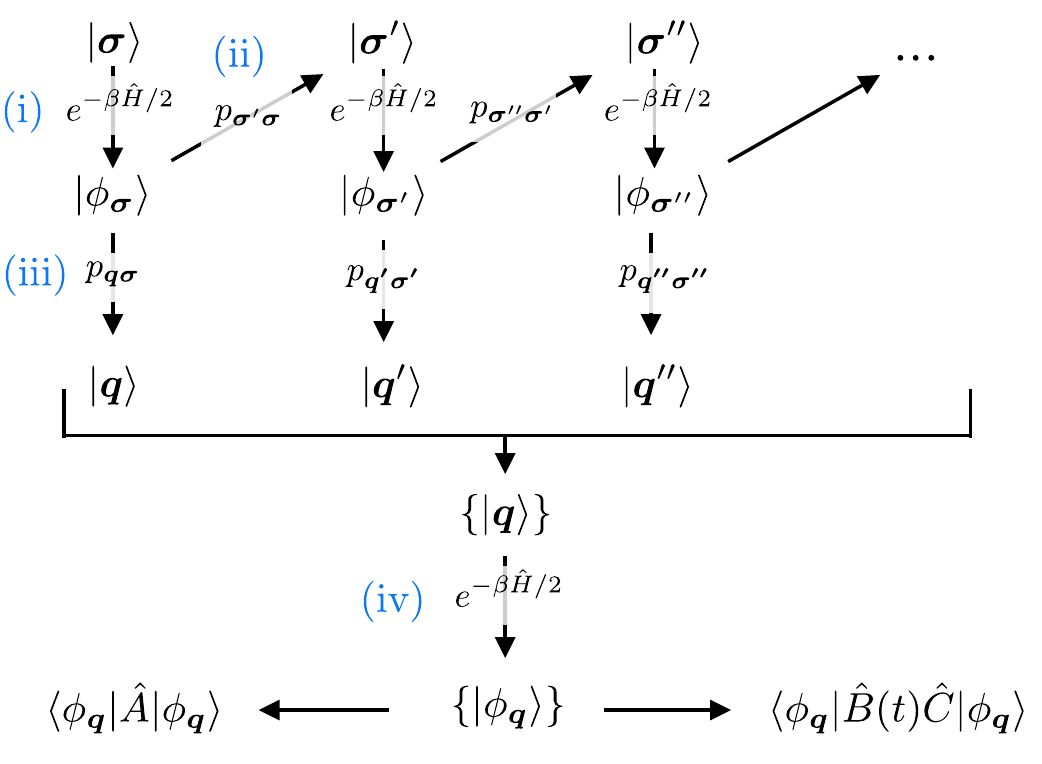}
\vspace{-10pt}
\caption{Schematic illustration of the SYMETTS sampling algorithm. For details on how to explicitly evaluate response functions of the type $\langle \hat{B}(t) \hat{C} \rangle_{\beta}$ see \Sec{sec:dyn}.}
\label{fig:SYMETTS_alg}
\end{figure}

\begin{compactenum}[(i)]
\setcounter{enumi}{2} 
 \item Using a symmetry-conserving collapse routine (described below), we collapse $|\phi_{\vsigma} \rangle$ to a SPS $|\vq\rangle$ with probability $p_{\vq \vsigma} = | \langle \vq | \phi_{\vsigma}\rangle|^2$. Each collapse generates a SPS according to the correct probability distribution $P_{\vq}/Z_{\beta}$ (thus belonging to one of the relevant symmetry sectors at a given temperature), as long as the non-symmetric METTS has been sampled according to $P_{\vsigma}/Z_{\beta}$. 
 \item The resulting SPS $ |\vq\rangle $ can easily be converted into an MPS with explicit encoded symmetry sectors,\cite{weichselbaum12a} which is then evolved in imaginary time and normalized to generate the SYMETTS  $|\phi_{\vq}  \rangle$. \\
\end{compactenum}
The combination of thermal and symmetrization step is then repeated with the newly generated CPS $|\vsigma '\rangle$ to create a full SYMETTS sample $\{ |\phi_{\vq}  \rangle\}$, which represents the basis for calculating static or dynamic observables at finite temperature (see \Fig{fig:SYMETTS_alg} for an illustration). Thus we ensure that all computed SYMETTS are minimally autocorrelated, as each of them is generated from a different non-symmetric METTS. 

By maximizing the ergodicity of the sample, we face additional computational cost, as we have to generate a full non-symmetric METTS sample $\{|\phi_{\vsigma} \rangle\}$ as well. In principle, it is possible to reduce the number of non-symmetric METTS by generating a larger number of SPS from each $|\phi_{\vsigma} \rangle$ by repeating the symmetrization step multiple times. However, this may introduce artificial correlations between different SPS generated from the same non-symmetric METTS. For this reason, we present the SYMETTS algorithm using a formulation that maximizes ergodicity. This limits the applicability of SYMETTS in terms of calculating static observables. In these cases, we would have to work harder than with regular METTS. However, when calculating \emph{dynamic} quantities, generating the SYMETTS sample accounts for only a factor $\mathcal{O}(10^{-3})$ or less of the total computation time. Hence, our algorithm ensures that the full potential of SYMETTS towards dynamic applications is guaranteed and no ergodicity problems arise.

\subsection{Collapse routine for non-Abelian symmetries}
 In this section, we illustrate step (iii) of the SYMETTS sampling for the example of the isotropic spin-$\frac{1}{2}$ Heisenberg chain (\ref{eq:XXX_H}). 

In  context of U$(1)_{\rm spin}$,  the collapse routine employed in step (iii) simply corresponds to the $z$ collapse discussed in \Sec{sec:ergo}, i.e., measuring along the $z$ axis only. The resulting SPS take the form of direct products of local symmetry eigenstate $|\uparrow\rangle, | \downarrow\rangle$ and are automatically distributed  according to the correct probability $P_{\vq}/Z_{\beta}$.
 
However, to exploit the full SU$(2)_{\rm spin}$ symmetry of the model, the collapse routine has to be adapted in order to generate a SPS ensemble $\{ |\vq \rangle \}$ of SU(2) eigenstates. For a single SPS, this is achieved by using a non-symmetric METTS $|\phi_{\vsigma} \rangle$ and sequentially collapsing it into the different eigensectors of the total spin operator $\hat{\vec{S}}^2$. 

To this end, we gradually build $\hat{\vec{S}}^2$ starting from the left end of the chain. At the first site, the total spin is always $\frac{1}{2}$ as we only consider a single spin, hence after constructing  $\hat{\vec{S}}_1^2$  no projection is required and the orthonormal center of $|\phi_{\vsigma} \rangle$ can be shifted to the second site. Here, we generate the total spin operator of first and second site according to
\begin{equation}
\hat{\vec{S}}^2_{L,2} = \hat{\vec{S}}_{1}^2 + \hat{\vec{S}}_2^2 + \hat{\vec{S}}_{1} \hat{\vec{S}}_2 + \hat{\vec{S}}_2 \hat{\vec{S}}_{1},
\end{equation}
with $\hat{\vec{S}}_j^2 = (\hat{S}^x_j)^2 + (\hat{S}^y_j)^2 + (\hat{S}^z_j)^2$ and  the subscript ``$L,2$'' indicating that we consider the total spin of the left part of the chain up to the second site. Diagonalizing this operator, we obtain the two spin sectors $S_{L,2}= \frac{1}{2} \pm \frac{1}{2} = 0, 1$ corresponding to the singlet and triplet configuration, and the projectors $\hat{P}(S_{L,2})$. We project the second bond of $|\phi_{\vsigma}  \rangle$ (and also $\hat{\vec{S}}^2_{L,2}$) either into singlet or triplet configuration according to the transition probabilities
\begin{equation}
p(S^2_{L,2} ) = \langle \phi_{\vsigma} | \hat{P}(S^2_{L,2}) |\phi_{\vsigma}  \rangle,
\end{equation}
and shift the orthonormal center of $|\phi_{\vsigma} \rangle$ to the next site. This procedure is repeated sequentially for every site $j$ of the system. Each time, we construct the spin operator for the left and the local part of the chain according to
\begin{equation}
\hat{\vec{S}}^2_{L,j} = \hat{\vec{S}}_{L,j-1}^2 + \hat{\vec{S}}_j^2 + \hat{\vec{S}}_{L,j-1} \hat{\vec{S}}_j + \hat{\vec{S}}_j \hat{\vec{S}}_{L,j-1}, 
\end{equation}
where  $\hat{\vec{S}}_{L,j-1}^2$ denotes the total spin squared of all sites to the left of (and excluding) site $j$. After diagonalization, the transition probabilities are calculated and $|\phi_{\vsigma}  \rangle$ is projected at bond $j$  into a single spin sector. Just as for the initially considered example, the operator $\hat{\vec{S}}^2_{L,j}$ always contains only two spin sectors, namely, $S_{L,j} = S_{L,j-1} \pm \frac{1}{2}$. Hence, diagonalization and projections can be carried out very efficiently.

In the end, one obtains a SU$(2)_{\rm spin}$ symmetric SPS $|\vq\rangle$ with probability $p_{\vq \vsigma} = | \langle \vq | \phi_{\vsigma}\rangle|^2$. States of this type are the initial point for setting up the SU(2) symmetric MPS framework.\cite{weichselbaum12a}


\section{Benchmark results} \label{sec:results1}
In this section, we present some benchmark results for our SYMETTS approach applied to both static and dynamic observables of the XXZ Heisenberg model with $N=100$ spins in the isotropic ($\Delta=1$, XXX model) and the free fermion limit ($\Delta=0$, XX model). As truncation criterion, we choose to keep all singular values above $s_{\beta}^{\rm tol} > 10^{-5}$ during the process of imaginary-time evolution, which is carried out using standard tDMRG tools with a second order Trotter decomposition and a time-step $\tau = 0.05 $. For the subsequent real-time evolution we adapt only the truncation criterion to $s^{\rm tol}_{\rm dyn} =  10^{-4}$. All quantities are expressed in terms of the coupling $J=1$.

\begin{figure}[t]
\centering
\includegraphics[width=\linewidth]{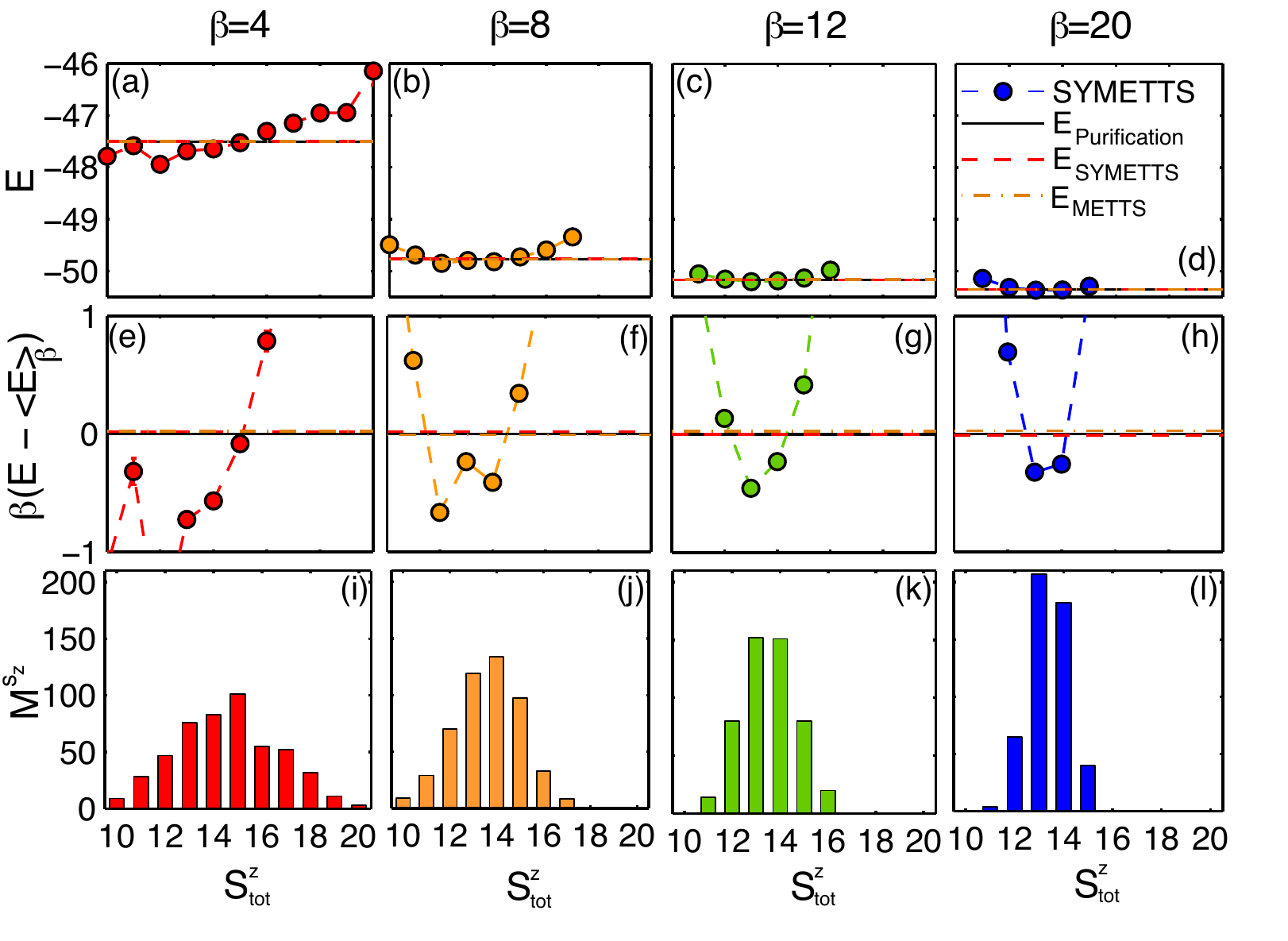}
\vspace{-10pt}
\caption{U(1)-SYMETTS sampling of the thermal energy for the isotropic Heisenberg chain with $h=1$. Panels (a-d) in the upper row display the thermal energy of each symmetry sector entering the SYMETTS sample for $\beta$ ranging between $4$ and $20$; dashed lines indicate the overall ensemble average $\langle E \rangle_{\beta}$ determined by SYMETTS. Moreover, a comparison to benchmark calculation based on METTS (dotted lines) and purification (solid lines) is provided. Panels (e-h) in the second row show $\beta(E-\langle E \rangle_{\beta})$, in order to zoom into an energy window of order of the temperature around $\langle E \rangle_{\beta}$. Panels (i-l) in the last row illustrates the subsample size $M^{S^z_{\rm tot}}$ of different symmetry sectors for a fixed total sampling size $M=500$.}
\label{fig:HB_N50U1hz1_M500}
\end{figure}

\subsection{Static observables: Thermal energy}

First, we discuss some static SYMETTS calculations for the total energy of the isotropic Heisenberg chain (\ref{eq:XXX_H}) with and without finite magnetic field. The data below conclusively show that the slightly modified METTS algorithm above is able to obtain results of similar accuracy as the non-symmetric METTS sampling at equal sample size $M$. Of course, this is to be expected since SYMETTS essentially generates the ensemble states analogously to the original algorithm. Nevertheless, this exercise helps to understand the importance of using sample states which are correctly distributed over the relevant symmetry sectors.

To illustrate the method in more detail, \Fig{fig:HB_N50U1hz1_M500} shows U(1)-SYMETTS results resolving the different symmetry sectors entering into the calculation of the thermal energy for four different inverse temperatures. The upper row displays the average energy of each subsample of states characterized by fixed $S^z_{\rm tot}$.  In the middle row, we zoom into a window of order of the temperature around the average energy of the sample. The resulting values for $\langle E \rangle_{\beta}$ determined by SYMETTS [\Eq{eq:expT_SPS}] are benchmarked against METTS calculations [\Eq{eq:expT}] and quasi-exact purification data. The thermal average of all SYMETTS subsamples leads to highly accurate results for $\langle E \rangle_{\beta}$.

For large $\beta$, the lowest energy is obtained by the $S^z_{\rm tot} = 12$ sector, which corresponds to the ground state symmetry sector of the system. Depending on the temperature, the energy of the neighboring sectors increases more or less steeply. At high temperatures, thermal fluctuations become clearly visible in the thermal energies of the different symmetry sectors. Accordingly, the number of relevant symmetry sectors obtained from the SYMETTS sampling step (ii) is closely related to the temperature. Thermal fluctuations drive the sample states into more ``excited'' symmetry sectors at high temperatures: the maximum symmetry sector occurring in the $\beta=4$ simulation corresponds to $S^z_{\rm tot}=20$, whereas we find a maximum of $S^z_{\rm tot}=15$ for $\beta=20$ as the system relaxes more towards the ground state. 
 This behavior is  also illustrated by the bottom row of \Fig{fig:HB_N50U1hz1_M500}, which shows the subsample size $M^{S^z_{\rm tot}}$ of each symmetry sector for a fixed total sample size $M = \sum_{S^z_{\rm tot}} M^{S^z_{\rm tot}} = 500$.  Again we observe, that the distribution of symmetry sectors is broad at high temperatures and becomes narrow for large values of $\beta$.    
 
 \begin{figure}[t]
\centering
\includegraphics[width=\linewidth]{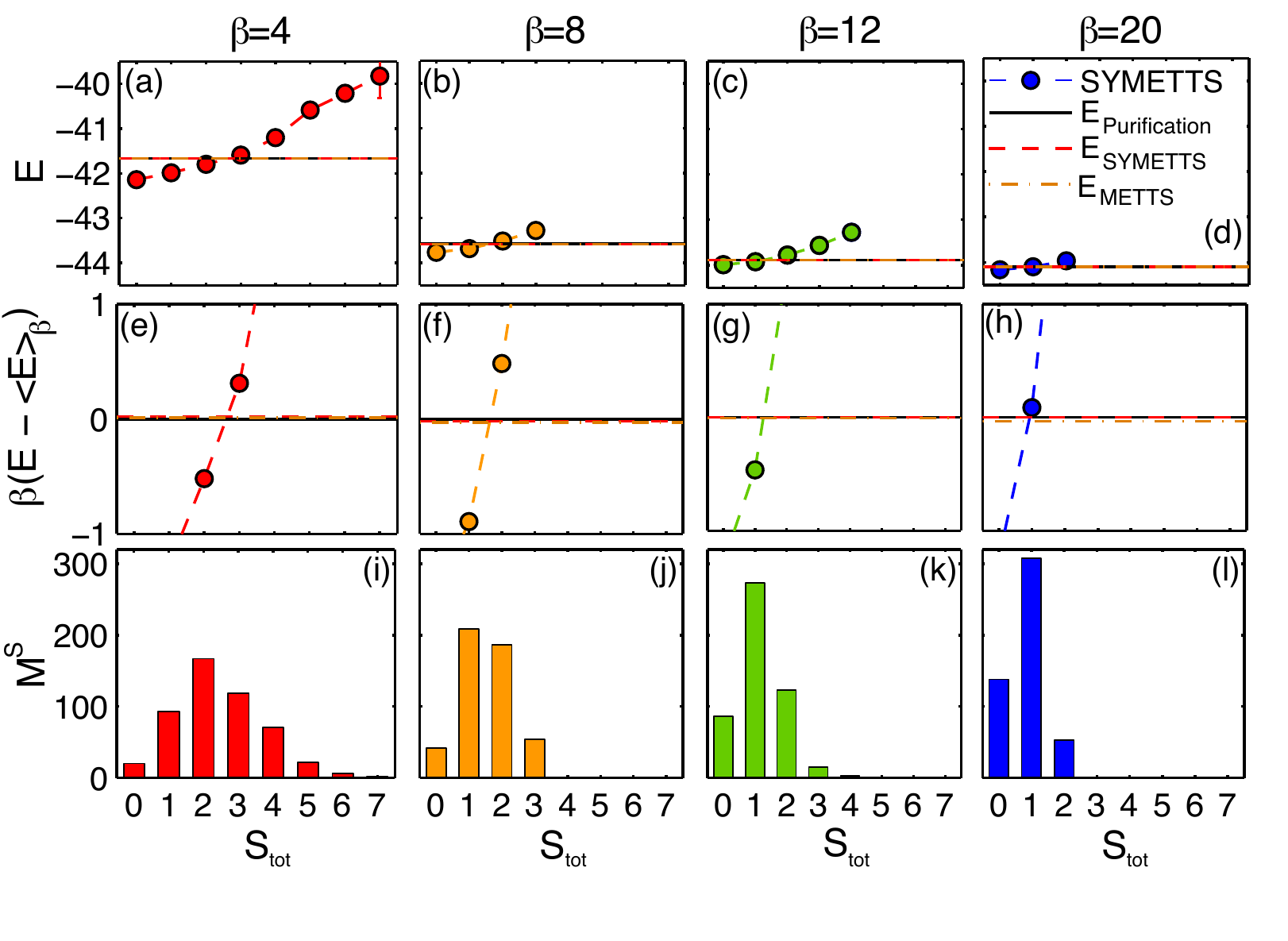}
\vspace{-15pt}
\caption{SU(2)-SYMETTS sampling of the thermal energy for the isotropic Heisenberg chain with $h=0$, using the same layout as \Fig{fig:HB_N50U1hz1_M500}.}
\label{fig:HB_N50SU2}
\end{figure}

Figure \ref{fig:HB_N50SU2} presents results for the thermal energy of the isotropic Heisenberg chain (\ref{eq:XXX_H}) at zero magnetic field, where we can exploit the non-Abelian SU(2) symmetry of the model. The layout of the panels and the parameters are chosen in accordance with those in  \Fig{fig:HB_N50U1hz1_M500} above. Instead of $S^z_{\rm tot}$ sectors, the SYMETTS are now categorized in terms of the total spin $S_{\rm tot}$ of each SU(2) multiplet in the sample. Again, the upper and middle rows display the average energy of each subsample corresponding to a fixed $S_{\rm tot}$. The resulting values for $\langle E \rangle_{\beta}$ determined by SYMETTS [\Eq{eq:expT_SPS}] are benchmarked against METTS calculations [\Eq{eq:expT}] and quasi-exact purification data. We find that the overall ensemble average of all SYMETTS subsamples gives a good approximation of the thermal energy of the state also for the non-Abelian sampling routine.

As for the Abelian case, the distribution of different multiplets shown in the last row becomes more narrow towards lower temperatures. Whereas the majority of states at $\beta=20$ belong to the multiplets $S_{\rm tot} = 0,1$, these sectors deplete for higher temperatures and the maximum moves towards $S_{\rm tot} = 2$ for $\beta=4$.

 \begin{figure}[b]
         \centering
\includegraphics[width=\linewidth]{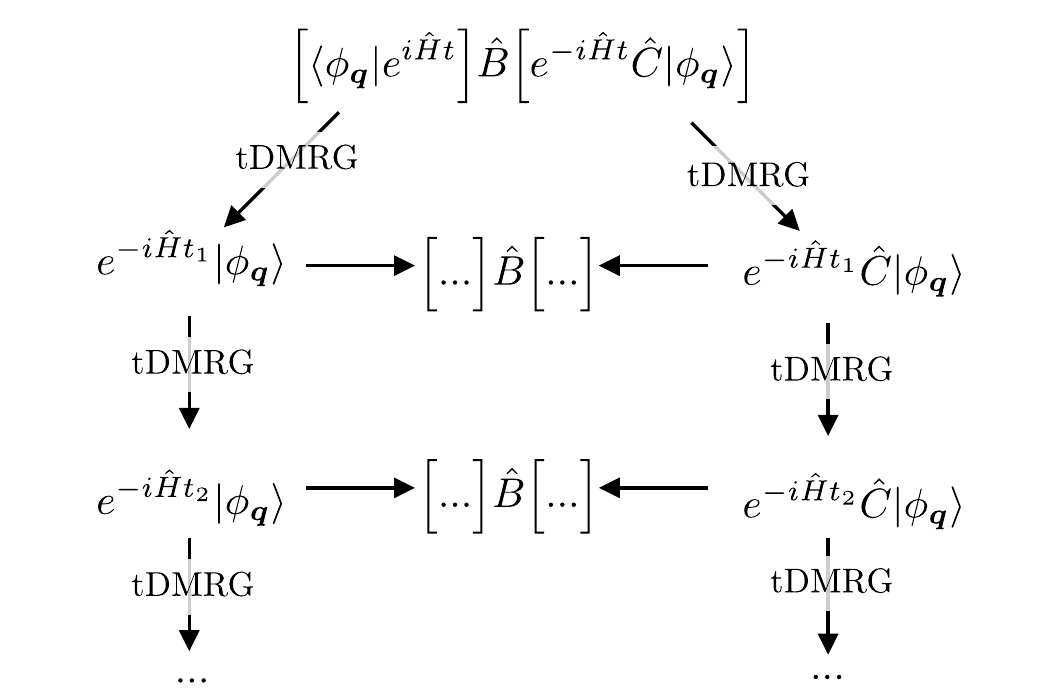}
\vspace{-10pt}
\caption{Schematic illustration of the dynamic calculation with SYMETTS.\cite{binder2014minimally}}
\label{fig:SYMETTS_dyn}
\end{figure}

\subsection{Dynamic observables: Dynamic spin structure factor  } \label{sec:dyn}
Whereas SYMETTS does not offer any significant computational advantage over the original formulation for computing static observables, its potential is enormous for the calculation of dynamic quantities, such as response functions of the form
\begin{equation} \label{eq:response}
\mathcal{A}_{\beta}^{\hat{B}\hat{C}}(t) = \langle \hat{B}(t)\hat{C} \rangle_{\beta}\ , \quad \text{with}\  \hat{B}(t) = e^{i\hat{H}t} \hat{B} e^{-i\hat{H}t}.
\end{equation} 
 For such problems, generating the ensemble states represents only a negligible part of the total computational costs. Most computational effort has to be put into the real-time evolution of each state in the sample, as the linearly growing entanglement requires an exponential increase of the bond dimension of the MPS towards longer time scales. Here SYMETTS offers a great advantage over the existing METTS approach, since the symmetry implementation strongly increases the numerical efficiency during the real-time evolution. In addition, the overhead cost of generating both a symmetric and non-symmetric sample in the SYMETTS sampling (see \Sec{sec:sampling}) can be ignored in almost every case, as it only accounts for a very small fraction [$\mathcal{O}(10^{-3})$] of the total computational time. The achievable efficiency gains are completely analogous to the exploitation of symmetries in other MPS applications, such as ground state DMRG, tDMRG or iTEBD. For example, the direct implementation of the Abelian U(1) symmetries in spin models can already speed up calculations by about a factor of up to $10$.\cite{Singh_PRB_2012,Kennes_2014}\textsuperscript{,}\footnote{See also App.~\ref{app:cputime}} Even larger benefits can be achieved when studying models with multiple Abelian or non-Abelian symmetries.

To simulate a response function using real-time evolution, we follow \Ref{binder2014minimally} and compute for every SYMETTS $|\phi_{\vq} \rangle$ in our sample the expectation value
\begin{equation}
\label{eq:Resp_METTS}
[  \langle  |\phi_{\vq}  | e^{i\hat{H}t} ] \hat{B}  [ e^{-i\hat{H}t}  \hat{C} |\phi_{\vq} \rangle] 
\end{equation}
by carrying out two independent real-time evolutions, $|\psi_{\vq}(t) \rangle = e^{-i\hat{H}t}  \hat{C} |\phi_{\vq} \rangle$ and $ |\phi_{\vq} (t)\rangle=e^{-i\hat{H}t}  |\phi_{\vq} \rangle$ using standard tDMRG. Equation \ref{eq:Resp_METTS} can be evaluated at any intermediate time-step $t$ by calculating the overlap for the operator $\langle \psi_{\vq}(t) | \hat{B}|\phi_{\vq} (t)\rangle$, as illustrated in \Fig{fig:SYMETTS_dyn}. In the end, we take the sample average to obtain a result for the finite-temperature response function.

In principle, there are other options for calculating the real-time evolution of response functions.\cite{binder2014minimally} In this work, we restricted ourselves to the scheme outlined above, as it requires only two tDMRG simulations per sample state to access all intermediate time-steps up to the maximally reached timescale $t_{\rm max}$. 

Instead of studying real-time response functions, here we consider their Fourier transforms, i.e., spectral functions. More particularly, we focus on dynamic spin structure factors $S^{\alpha \beta} (\omega,k)$, which are the Fourier transform of dynamical spin correlation functions. These quantities are of particular experimental relevance, as they can be directly accessed by inelastic neutron scattering experiment. For a benchmark, we compute the dynamic spin structure factor of the XXZ Heisenberg model with open boundaries:
\begin{equation} \label{eq:DSSF_XX}
S^{\alpha \beta} (k, \omega) = \sum_{ij}^N \frac{\sin{(ik)} \sin{(jk)}}{\pi(N+1)}  \int {\rm d}t\ e^{i \omega t } \langle S^{\alpha}_i(t) S^{\beta}_j(0) \rangle. 
\end{equation}

To this end, we define the spin-wave operator $\hat{S}^{\alpha}_k = \sqrt{\frac{2}{N+1}} \sum_{j=1}^N \sin{\big( \frac{j \pi k}{N+1}  \big)} \hat{S}^{\alpha}_j$ and evaluate $\langle \hat{S}^{\alpha}_k(t) \hat{S}^{\beta}_k\rangle_{\beta}$ via \Eq{eq:Resp_METTS} for a number of intermediate points up to some maximum time $t_{\rm max}$. Then, we perform a Fourier transform to frequency space, including a Gaussian broadening $\exp[-4 (t/t_{\rm max})^2]$ in the integral in \Eq{eq:DSSF_XX} to remove artificial oscillations, which are caused by the finite cut-off of the real-time evolution.\cite{WhiteFeiguin04} This means that the exact spectral features are convolved with a Gaussian $\exp[ -\omega^2/(2 W^2)]$, with a frequency resolution $W=2\sqrt{2}t_{\rm max}^{-1}$. In some cases, linear prediction can be used to avoid the artificial broadening and extract more spectral information from the time series.\cite{White_PRB_2008_LinPre,Barthel_PRB_2009} However, we found that linear prediction is not reliable in our study of the generalized diamond chain (see \Sec{sec:Az_DSSF}). Hence, we refrain from employing linear prediction in this work. 

In a first study, we employ our U(1)-SYMETTS approach to extract the dynamic spin structure factor in the limit of $\Delta=0$. In this case, the XXZ model can be solved exactly by mapping the system by a Jordan-Wigner transformation to non-interacting spinless fermions.\cite{Stolze_PRB_1995,Derzhko_PSS_1998} This allows us to exactly evaluate the spin correlation functions $ \langle \hat{S}^{\alpha}_i(t) \hat{S}^{\beta}_j(0) \rangle$ for arbitrary times and obtain the dynamic spin structure factor by Fourier transformation for direct comparison to the SYMETTS data. 

\begin{figure}[t]
\centering
\includegraphics[width=\linewidth]{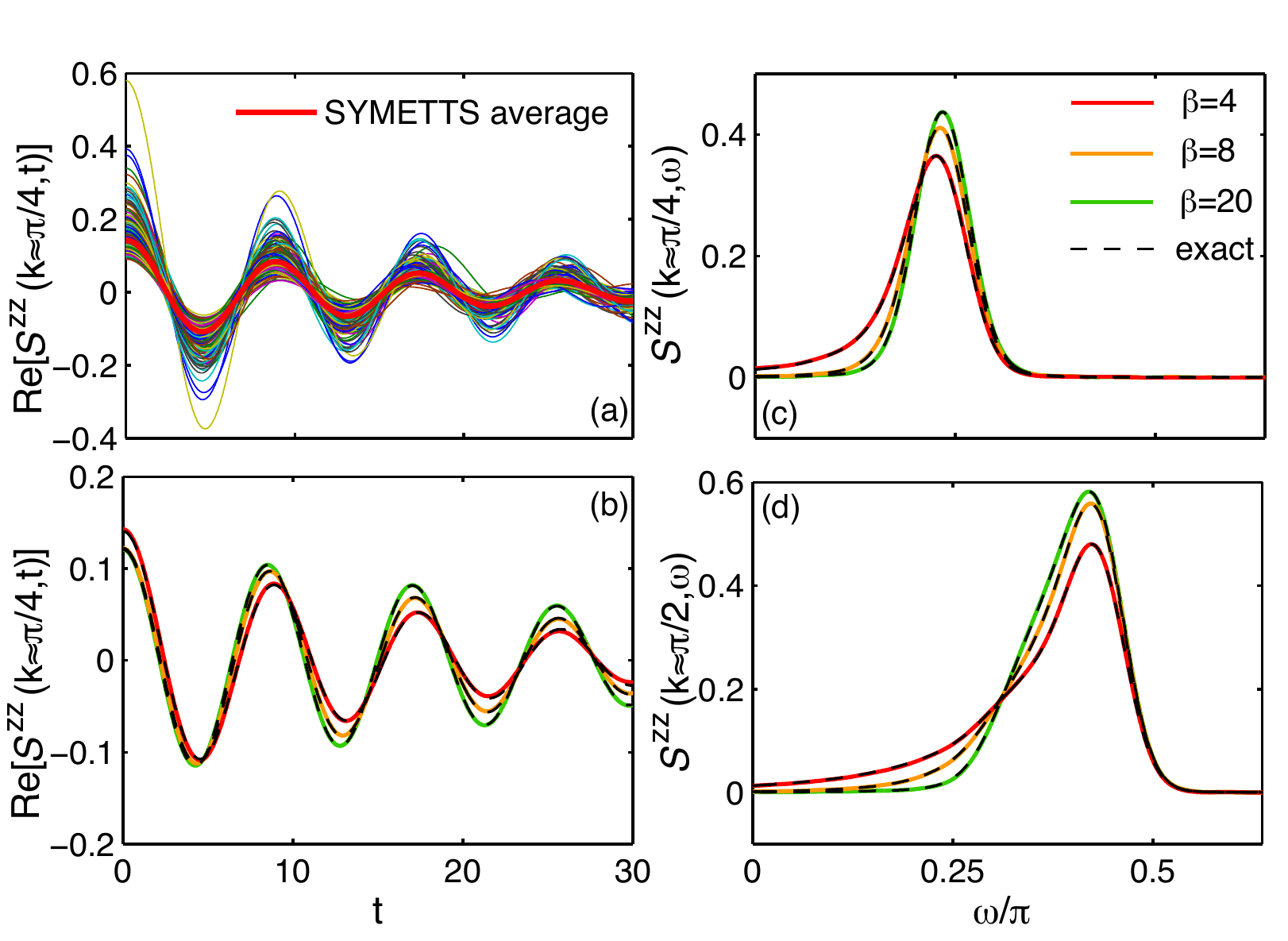}
\vspace{-10pt}
\caption{tDMRG calculation for the dynamic spin structure factor of the XX model using $M=300$ ensemble states. Panel (a) illustrates the time evolution of individual SYMETTS (thin lines) at $\beta=4$ used to calculate $S^{zz}(\pi/4,t)$ by taking the ensemble average (thick red line). (b) Displays the SYMETTS ensemble average for various inverse temperatures, which is then used to compute the dynamic spin structure factor in frequency space. Panels (c) and (d) show the frequency data obtained from Fourier transform for $k\approx \pi/4$ and $\pi/2$. For all considered inverse temperatures, we find excellent agreement with the exact result (dashed lines).}
\label{fig:HB_XX_DSSF_tDMRG}
\end{figure}

\begin{figure}[b]
\centering
\includegraphics[width=\linewidth]{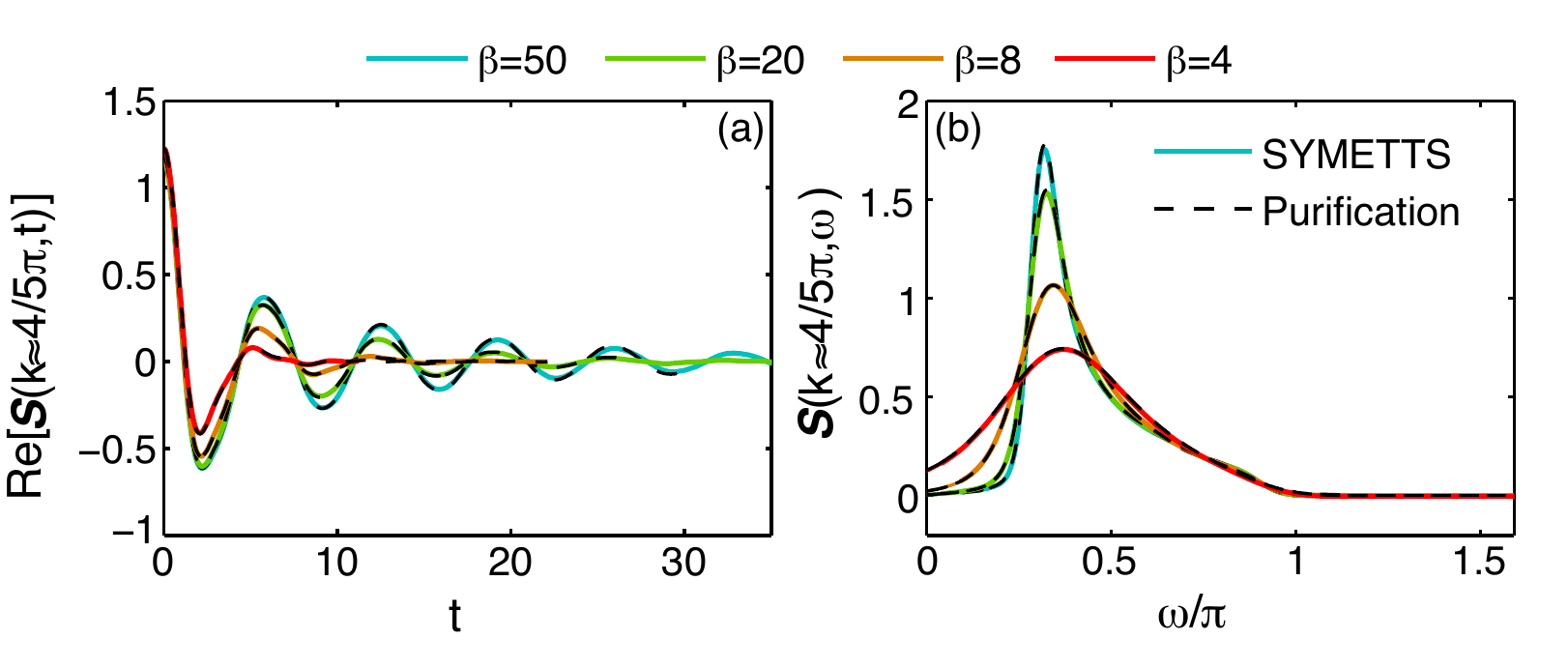}
\vspace{-10pt}
\caption{SU(2)-SYMETTS calculation (solid lines) for (a) $\vspin(3\pi/4,t)$ and (b) $\vspin(3\pi/4,\omega)$  of the isotropic Heisenberg chain using $M=300$ ensemble states. For all considered inverse temperatures, we find very good agreement with data obtained from matrix-product purification (dashed lines).}

\label{fig:HB_XXX_DSSF_tDMRG}
\end{figure}
\Fig{fig:HB_XX_DSSF_tDMRG}(a) displays the real-time evolution of $S^{zz}(\pi/4,t)$ for $\beta=4$ up to $t_{\rm max}=30$, with the thin lines corresponding to individual realization of particular SYMETTS states and the thick red line denoting the ensemble average. The sample averages are collected for different temperatures in \Fig{fig:HB_XX_DSSF_tDMRG}(b). After the real-time evolution, we perform a Fourier transform to obtain the dynamic spin structure factor as a function of frequency, as shown in  \Figs{fig:HB_XX_DSSF_tDMRG}(c,d) for  $k\approx \pi/2$  and $\pi/4$. We find excellent agreement with exact results (dashed lines). A prerequisite for agreement of this quality is that the statistical sampling error, and hence the temperature, is sufficiently small; for the sample size of $M=300$ used in \Fig{fig:HB_XX_DSSF_tDMRG} the relative error, defined as
\begin{equation}
\delta S = \frac{\sqrt{\int {\rm d} \omega [S(k,\omega) - S_{\rm exact}(k,\omega)]^2   } }{\sqrt{\int {\rm d} \omega S_{\rm exact}(k,\omega)^2 }},
\end{equation}
varies between $\delta S \approx 1\%$ for $\beta=4$ and $\delta S \approx 0.3\%$ for $\beta=20$. We note that the error is approximately proportional to the temperature, which indicates that the dominant contribution is given by the statistical error of the ensemble, which scales as $\sim T/\sqrt{M}$. 

\begin{figure}[t]
\centering
\includegraphics[width=\linewidth]{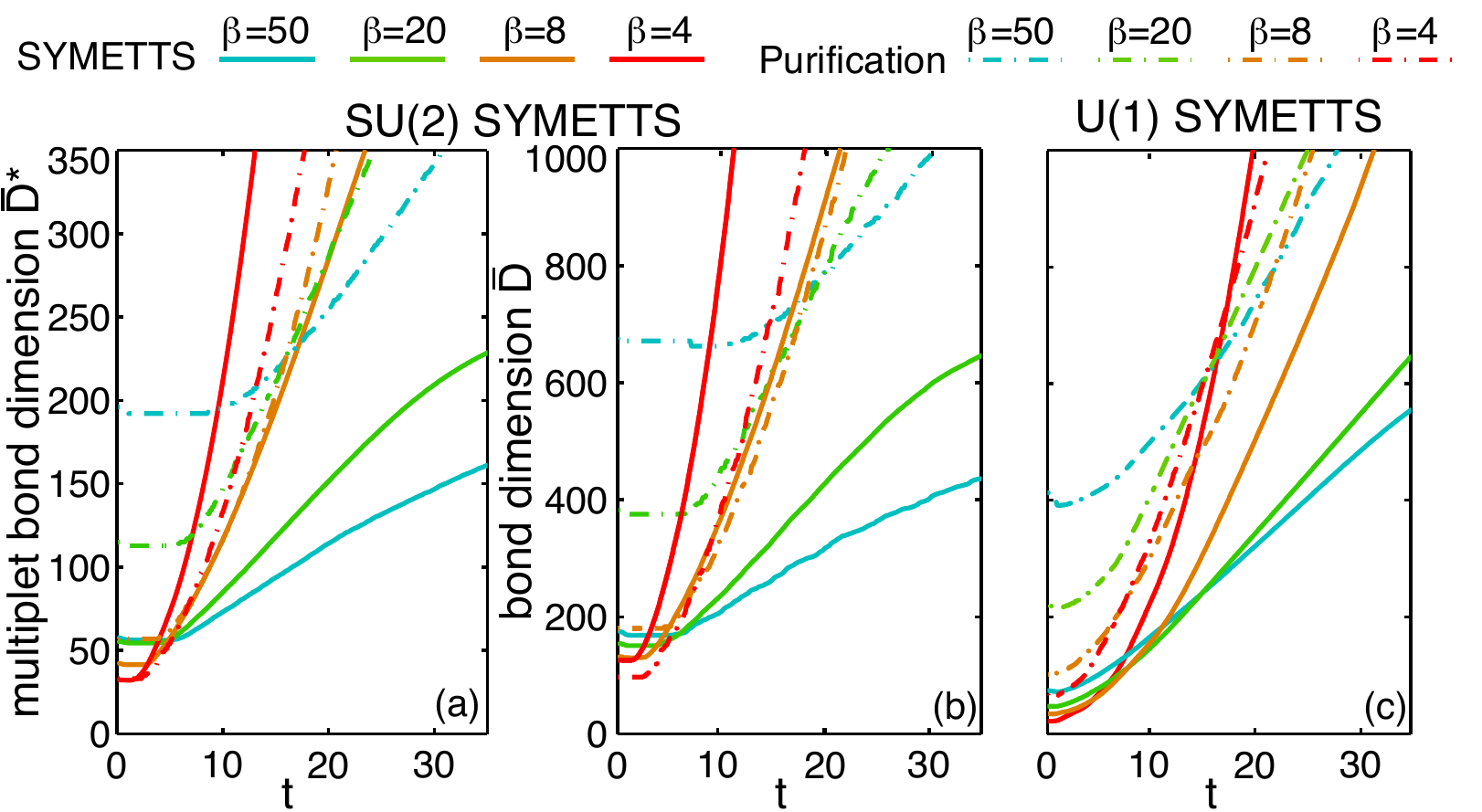}
\vspace{-10pt}
\caption{(a) Average multiplet bond dimension $\bar{D}^*$ and (b) the corresponding bond dimension $\bar{D}$ of SU(2)-SYMETTS during the calculation of $\vspin(3\pi/4,t)$ (solid lines) in comparison to  the purified density matrix (dashed lines). In both cases, we keep all singular values $>10^{-4}$ during real-time evolution. For $\beta=4,8$ we include a backwards time-evolution of the auxiliary bonds of the purified density matrix, as this leads to a reduction of the entanglement growth.\cite{Karrasch_PRL_2012,Karrasch_NJP_2013} (c) Average bond dimension $\bar{D}$ of U(1)-SYMETTS sample during the calculation of $S^{zz}(3\pi/4,t)$ for comparison. }
\label{fig:HB_XXX_DSSF_D}
\end{figure}

Next, we consider an isotropic coupling $\Delta=1$, which allows us to compute $\vspin(k,\omega)$ using SU(2)-SYMETTS, since the XXZ Hamiltonian (\ref{eq:XXX_H}) features the full spin symmetry in this case. Figures  \ref{fig:HB_XXX_DSSF_tDMRG}(a) and \ref{fig:HB_XXX_DSSF_D}(b) show the results for $k\approx3\pi/4$ in time and frequency space, respectively, in comparison to purification calculations using the same truncation criterion (black dashed lines). The maximum time $t_{\rm max}$ varies for different temperatures, since we stopped the SYMETTS calculations when a threshold of 1000 states was exceeded by the bond dimension $D$, which was determined adaptively by keeping all singular values $>10^{-4}$ (see \Fig{fig:HB_XXX_DSSF_D} and upper panel of Table \ref{tableD} for values of $t_{\rm max}$). Again, we find excellent agreement for the considered temperature range.

In this context, we briefly discuss the intriguing question whether SYMETTS can reach longer time scales than purification in certain limits. To this end, we study the growth of entanglement during the real-time evolution, which manifests itself in the growing bond dimension of both the average SYMETTS as well as the purified density matrix. Figures \ref{fig:HB_XXX_DSSF_D}(a) and \ref{fig:HB_XXX_DSSF_D}(b) present SU(2) data for the average maximum multiplet bond dimension $\bar{D}^*$ and the corresponding states space dimension $\bar{D}$, respectively. We find that an average SYMETTS requires significantly less numerical resources at $\beta=20,50$. For such low temperatures, SYMETTS certainly allows to access longer time scales than purification when fixing the numerically feasible bond dimension to an upper cutoff. Note that due to the presence of the statistical error, this does not imply that SYMETTS is generally more accurate than purification when fixing the total computation time and judging accuracy by comparing to quasi-exact calculations, as done in \Ref{binder2014minimally}. However, if one does not insist to push the statistical error towards the order of the truncation error, and moreover takes into account parallelizability, SYMETTS offers much potential towards the dynamical description of low-dimensional systems at low temperature. This is already illustrated by the calculations in this sections, demonstrating that it is possible to extract  the dynamic structure factor with high accuracy using a sample size of only a few hundred states.

On the other hand, SYMETTS is limited to small $t_{\rm max}$ at high temperatures. Particularly at  $\beta=4$, a single SU(2)-SYMETTS on average requires larger bond dimensions $\bar{D}$ than the purified density matrix! This can be attributed to the intrinsic structure of the SU(2) symmetry product states. Although their multiplet dimension $D^*$ is strictly unity at infinite temperature, the SPS already contain some entanglement due to the presence of non-trivial multiplet sectors with internal structure, which lead to a state space dimension $D>1$. Because of thermal fluctuations, these ``excited'' multiplet sectors appear more frequently at high temperatures, which is illustrated by the comparison $\bar{D}^*$ and $\bar{D}$ of the corresponding SPS samples in the lower panel of Table \ref{tableD} for different temperatures. With these non-trivial multiplets being present in the SPS, the subsequent imaginary- and real-time evolution obviously also induces more entanglement. This explains why $\bar{D}$ of a SU(2)-SYMETTS exceeds the bond dimension necessary to represent the purified density matrix  already before starting the real-time evolution at $\beta=4$ [c.f~\Fig{fig:HB_XXX_DSSF_D}(b)].  

This issue is not present in the context of U(1)-SYMETTS, where the SPS does not contain any intrinsic entanglement at infinite temperature and thus can still be considered as a classical product state. Thus, the initial $\bar{D}$ is strictly smaller than the bond dimension of the purified density matrix for all temperatures, as shown in\Fig{fig:HB_XXX_DSSF_D}(c). Moreover, the increase of $\bar{D}$ at high and intermediate temperatures is slightly less severe than in the SU(2) calculations.  

We conclude from this analysis that it is possible for SYMETTS to exploit both Abelian and non-Abelian symmetries. In combination with tDMRG, it represents a valuable alternative to computing spectral functions, particularly for low temperatures. In addition, we find that Abelian SYMETTS have favorable entanglement properties over their non-Abelian counterparts at high temperatures. Hence, one should refrain from exploiting non-Abelian SYMETTS in these cases and switch to U(1)-SYMETTS or matrix-product purification.

\begin{table}[h!]
\begin{center}
\begin{tabular}{| r | c | c | c | c | }
\hline
 $t_{\rm max}$    & $ \beta=4 \phantom{-}$ & $\beta=8\phantom{-}$& $\beta=20$      &  $\beta=50$      \\
\hline
SYMETTS${}_{\rm SU(2)}$ & 11.2 & 21.4& $>35$ & $>35$ \\
Purification${}_{\rm SU(2)}$  & 18 & 22 & 26.1& 30.6\\
SYMETTS${}_{\rm U(1)}$ & 19.8 & 31.2 & $>35$ & $>35$  \\

\hline \hline

bond dim.   & $ \beta=4 \phantom{-}$ & $\beta=8\phantom{-}$& $\beta=20$      &  $\beta=50$      \\
\hline
$\phantom{{}^{2^2}} \bar{D}^*_{\rm SU(2)} $& 1 & 1 & 1 & 1 \\
$\bar{D}_{\rm SU(2)} $& 4.11 & 3.08 & 2.31 & 1.96 \\
$\text{max}[D_{\rm SU(2)}] $& 16 & 12 & 7 & 5 \\
$\bar{D}_{\rm U(1)}$ & 1 & 1 & 1 & 1 \\
\hline 
\end{tabular}
\end{center}
\caption{Upper panel: maximum time $t_{\rm max}$ reached in the simulations shown in \Figs{fig:HB_XXX_DSSF_tDMRG} and \ref{fig:HB_XXX_DSSF_D}. Lower panel: average multiplet bond dimension $\bar{D}^*$ and corresponding average and maximum bond dimension for SU(2) and U(1) symmetry product states for various temperatures.}
\label{tableD}
\end{table}

We point out that the METTS algorithm in principle  can also exploit time-translational invariance in order to reformulate the response function in terms of $\langle \hat{B}(t/2) \hat{C}(-t/2) \rangle$, which effectively doubles the maximum reachable time scale $t_{\rm max}$.\cite{barthel_2012scaling,Barthel_NJP_2013,binder2014minimally} Ideally, the time-evolution is then carried out in the Heisenberg picture by evolving $\hat{B}$ and $\hat{C}$ directly in terms of matrix-product-operators (MPO),\cite{Pizorn_NJP_2014} so that it still requires only two tDMRG simulations to access all intermediate time-steps. We note that working in the Heisenberg picture is generally considered to be suboptimal for matrix product purification.\cite{Barthel_NJP_2013} However, it seems more appealing for the METTS framework as one could carry out the real-time evolution only once for the MPO and compute the response function by calculating the overlap of the time-evolved MPO and the METTS sample. Thus the time-evolved MPO could be recycled for arbitrary temperatures. In general, this would imply that the maximum reachable time scale is set by the real-time evolution of the operators. In the pure-state formulation, $t_{\rm max}$ would then be temperature independent, as temperature only enters through the calculation of the overlaps with the METTS sample. Naturally, this idea enormously profits from the inclusion of symmetries into the METTS language presented here, but is beyond the scope of this paper and will be discussed elsewhere.

Finally, we remark that we have also explored the possibility of combining SYMETTS with a Chebyshev expansion to directly compute spectra in frequency space. However, this approach is computationally more expensive due to technical reasons and therefore not recommendable (see Appendix \ref{sec:CheMETTS} for details).


\FloatBarrier
\section{Generalized diamond chain model for azurite} \label{sec:azurite}
In the following, we demonstrate the efficiency of SYMETTS by studying a more complicated spin-chain model of direct experimental relevance. We focus on the
natural mineral azurite $\text{Cu}_3(\text{CO}_3)_2(\text{OH})_2$, which has attracted much attention due to the discovery of a plateau at $\frac{1}{3}$ in the magnetization curve at low temperatures.\cite{kikuchi2003magnetic,Kikuchi_PRL_2005,GU_PRLCom_2006,Gu_PRB_2007,Mikeska_PRB_2008,Rule_PRL_2008,Kang_JCMP_2009,jeschke2011multistep,Honecker_JCM_2011,rule2011dynamics,Ananikian_EPJB_2012,Cong_PRB_2014,Richter201539} Some authors proposed that the magnetic properties of this material are well described by a spin-$\frac{1}{2}$ diamond chain formed by the copper atoms with purely antiferromagnetic exchange couplings.\cite{Kikuchi_PRL_2005,Kikuchi_PRLReply_2006,Mikeska_PRB_2008} Others suggested a dominant ferromagnetic coupling\cite{GU_PRLCom_2006,Gu_PRB_2007,Rule_PRL_2008} and the importance of interchain coupling,\cite{Kang_JCMP_2009} yet none of them were able to derive a microscopic model for azurite that is able to fully characterize its complex magnetic properties.

\begin{figure}[t]
\centering
\includegraphics[width=\linewidth]{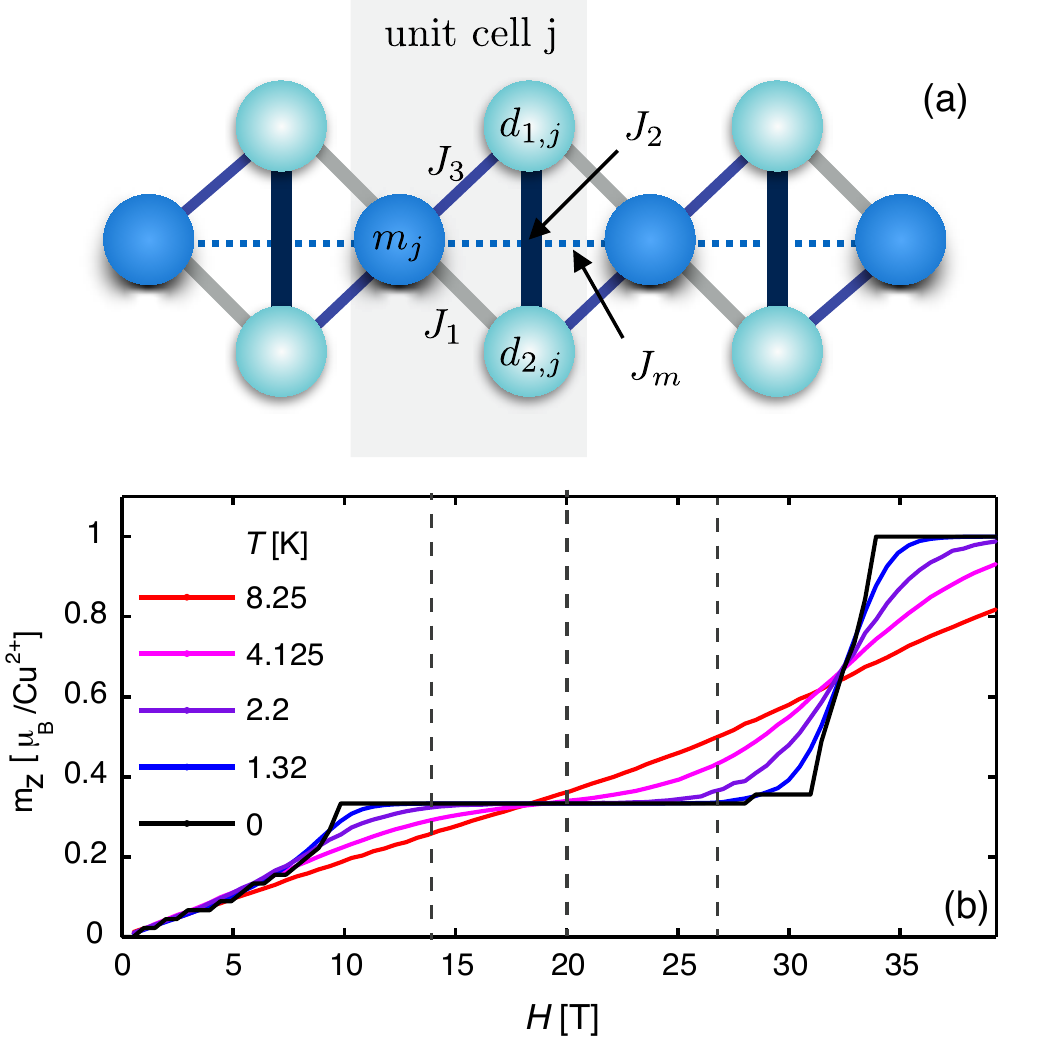}
\vspace{-10pt}
\caption{(a) Illustration of the generalized diamond chain model with the antiferromagnetic exchange couplings $J_1, J_2, J_3,$ and $J_m$. One unit cell of the system is highlighted by the gray area. (b) Dependence of the magnetization on an external magnetic field $H$, which we calculated by employing DMRG and SYMETTS at zero and finite temperature, respectively. In all calculations, we keep every singular value larger than the truncation threshold $s^{\rm tol}=10^{-5}$ and use a sample size of $M=1000$. For a system of $N=90$ spins in total, the emergence of the $\frac{1}{3}$ plateau can be observed for fields in the range of $H_{l,c}\leqslant H \leqslant H_{u,c}$. At finite temperatures, the plateau is washed out and the magnetization curve becomes a linear function of $H$. The vertical dashed lines indicate the parameter choices for our dynamical SYMETTS calculations in \Sec{sec:Az_DSSF}.}
\label{fig:azurite_plateau}
\end{figure}

Employing a combination of first-principle methods, exact diagonalization, and DMRG, \Ref{jeschke2011multistep} recently derived a full three-dimensional model which can be mapped to an effective one-dimensional system, namely a generalized diamond chain model with purely antiferromagnetic couplings, illustrated in \Fig{fig:azurite_plateau}(a). One third of the Cu spins (dark blue balls) forms weakly coupled monomers (dashed horizontal lines), whereas the other two thirds (light blue balls) form strongly coupled dimer singlets (heavy vertical lines). The dominant energy scale is determined by the dimer-dimer coupling $J_2$. In addition, there are nearest- and third-nearest neighbor dimer-monomer exchange $J_1$ and $J_3$ as well as the monomer-monomer coupling $J_m$. More precisely, the full Hamiltonian of the generalized diamond chain is defined as
\begin{eqnarray}
\hat{H}_0 &=& \sum_{j=1}^{N/3} \Big[ J_1 \vspin_{m,j} \cdot ( \vspin_{d1,j+1} + \vspin_{d2,j}) + J_2 \vspin_{d1,j} \cdot \vspin_{d2,j} \nonumber \\
 && + J_3 \vspin_{m,j} \cdot (\vspin_{d1,j} + \vspin_{d2,j+1}) + J_m \vspin_{m,j} \cdot \vspin_{m,j+1} \Big] \nonumber \\ 
 &&  -  g \mu_B H \sum_{j=1}^{N/3} \Big[ \hat{S}^z_{d1,j} + \hat{S}^z_{d2,j} + \hat{S}^z_{m,j} \Big],
\end{eqnarray}
with external magnetic field $H$, Bohr magneton $\mu_B$ and gyromagnetic ration $g=2.06$.\cite{Ohta_2003} $N$ labels the total number of Cu spins in the system,  the number of unit cells is therefore given by $N/3$. Note that this model features a U$(1)_{\rm spin}$ symmetry for finite values of $H$, which we exploit in our SYMETTS calculations. The value of the couplings has been determined by DFT calculations and small refinements using experimental data, leading to
\begin{eqnarray}
J_1 &=& 15.51\ {\rm K}, \quad J_2=33\ {\rm K},  \nonumber \\ 
 J_3 &=& 6.93\ {\rm K}, \quad J_m = 4.62\ {\rm K}.
\end{eqnarray}
Based on this system, the authors of \Ref{jeschke2011multistep} managed to derive a full microscopic picture for azurite, that is able to explain a wide number of experimental results. Additional support for the validity of this model is given by \Ref{Honecker_JCM_2011}, which explores further aspects such as magnetocaloric properties and excitation spectrum. Although Refs.~[\onlinecite{jeschke2011multistep}] and [\onlinecite{Honecker_JCM_2011}]  also present some selected results for the dynamic spin structure factor using dynamical DMRG (DDMRG), their resolution in the energy $\omega$ and the momentum transfer $k$ is limited, since DDMRG is numerically expensive and requires separate calculations for each $\omega$. Moreover, their results are restricted to zero temperature.

We will now illustrate the power of SYMETTS by calculating the excitation spectra in the plateau phase and analyze the influence of magnetic field and temperature on the excitation branches.

\begin{figure*}[t]
\centering
\includegraphics[width=\linewidth]{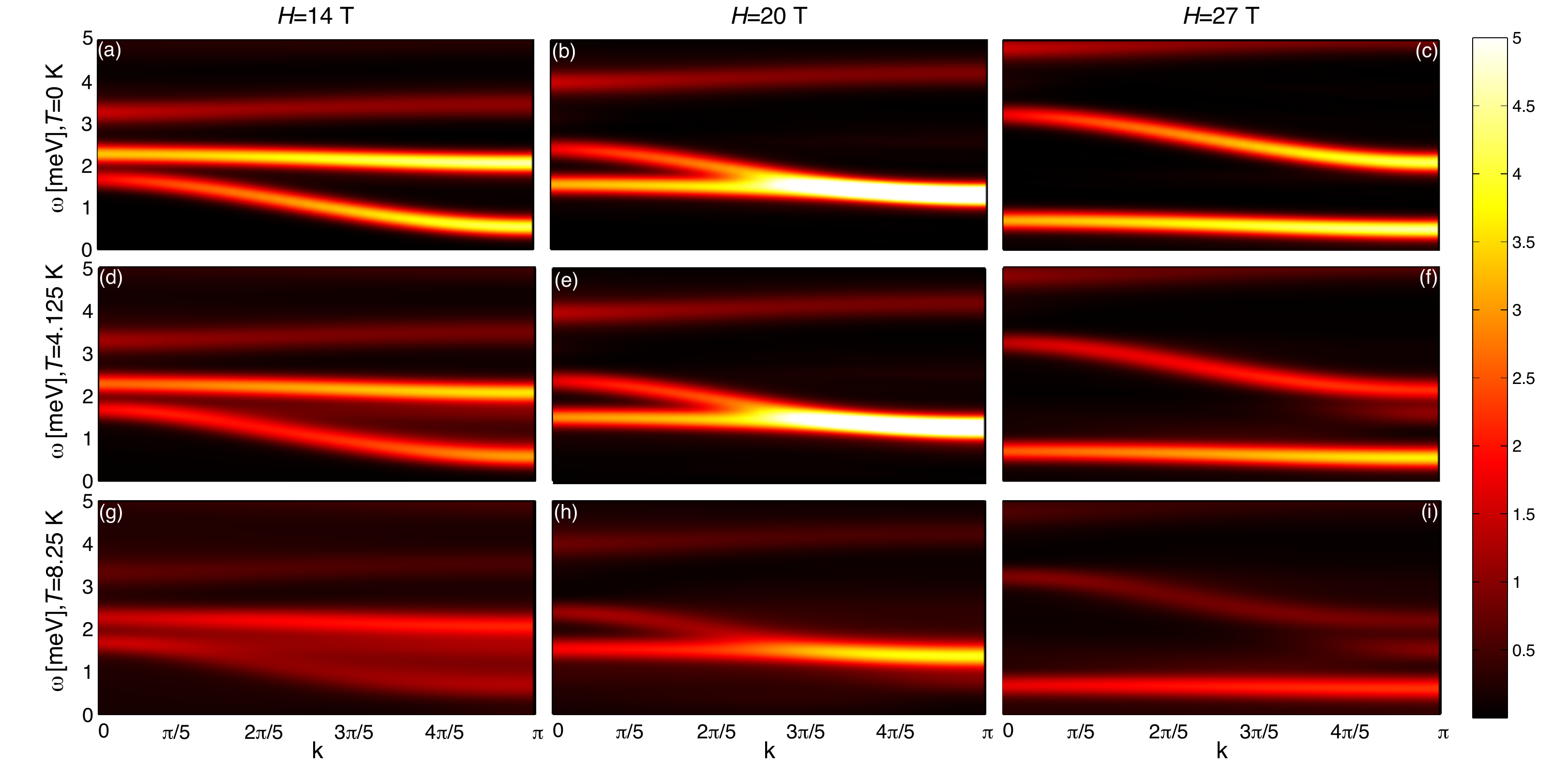}
\vspace{-15pt}
\caption{Transverse dynamic spin structure factor $S^{xx}(k,\omega)$ of a generalized diamond chain model for azurite. The intensity is displayed in arbitrary units. Each column indicates a different magnetic field strength, $H = 14, 20, 27$~T, corresponding to distinct points in the $\frac{1}{3}$ plateau phase. The ground state spectra are displayed in panels (a)--(c) in the first row, whereas the panels (d)--(i) show finite-temperature results obtained using tDMRG in combination with U(1)-SYMETTS (for details see text). }
\label{fig:DSSF_az}
\end{figure*}

\subsection{Magnetization plateau} \label{sec:plateau_az}
The most striking feature of azurite is the plateau at $\frac{1}{3}$ in the magnetization curve as a function of a magnetic field applied along the $b$ axis of the crystal lattice.\cite{Kikuchi_PRL_2005} This property can be nicely captured by the generalized diamond chain model, as already demonstrated in \Ref{jeschke2011multistep} by employing a direct comparison of the magnetization obtained from ground-state DMRG calculations with experimental data. Performing DMRG on an open chain with $N=90$ spins, we obtain the magnetization plateau at $T=0$ shown in \Fig{fig:azurite_plateau}(b) (black line). For better comparison, we use experimental units in the rest of this section. We find that the plateau phase is bounded by a lower and upper critical field, $H_{l,c} \approx 9.8\ {\rm T}$  and $H_{u,c} \approx 31.0\ {\rm T}$. Note that the small intermediate steps for $H<H_{l,c}$ and $H=28.5\ {\rm T}$ are artifacts caused by finite-size effects of the chain.

The plateau can be explained by a very intuitive argument.\cite{jeschke2011multistep} The dominant dimer-dimer exchange coupling $J_2$ forces the dimer spins into a singlet state, whereas the monomer spins are only weakly coupled by $J_m$. Therefore, the monomer spins polarize first for a finite magnetic field, whereas the dimers remain in the singlet state for a considerable interval of $H$. Only at large fields $H>H_{u,c}$, the dimers are arranged in a polarized state. Thus only $\frac{1}{3}$ of the total spins is aligned in direction of the field at intermediate fields strengths $H_{l,c}\leqslant H \leqslant H_{u,c}$, leading to the emergence of the $\frac{1}{3}$ magnetization  plateau.

Introducing thermal fluctuations by employing SYMETTS, we observe that the plateau is gradually washed out with increasing temperature. At high temperatures, the quantum mechanical properties of the system are almost erased and the magnetization curve becomes a linear function of the magnetic field. We note that in the plateau phase, the effect of temperature on the magnetization depends strongly on the specific field strength. For values of $H$ significantly smaller than $20$~T the magnetization strongly decreases with increasing temperatures. In this case, we expect the change in magnetization to be predominantly caused by the monomers, which have to vacate their fully polarized state. On the other hand, the monomers cannot contribute to the thermal increase of the total magnetization for $H>20$~T as they are already fully polarized on the plateau. Here the thermal fluctuations should predominantly excite the dimers by breaking up their singlet structure. We expect this to be reflected in the excitation spectra on the plateau. We study these next by means of the transverse dynamic spin structure factor, for the three values $H = 14, 20, 27$~T indicated by the vertical dashed lines in \Fig{fig:azurite_plateau}(b).

\subsection{Transverse dynamic spin structure factor} \label{sec:Az_DSSF}
We employ U(1)-SYMETTS to compute the transverse dynamic spin structure factor of the generalized diamond chain model, which can directly be measured by neutron scattering experiments.  Following \Ref{Honecker_JCM_2011}, the dynamic spin structure factor is defined as
\begin{equation} \label{eq:DSSF_az}
S^{xx}(\vec{k},\omega) = \frac{1}{N} \sum_{m,n} e^{i\vec{k}(\vec{R}_i - \vec{R}_j)}  \Big[ \int {\rm d}t \ e^{i \omega t} \langle \hat{S}^x_i (t) \hat{S}^x_j \rangle \Big].
\end{equation}
 Note that it is important to use the precise positions $\vec{R}_i $ of the Cu spins in azurite\cite{Zigan1972} and the experimentally chosen momentum direction in order to make the data comparable to the experiment in \Ref{Rule_PRL_2008}.

We perform all calculations for an open chain of $N=90$ spins, which allows an accurate resolution of the momentum transfer $k$ along the chain direction. For each $k$, we average over a SYMETTS sample of 300 states exploiting the U$(1)_{\rm spin}$ symmetry of the model. In comparison to non-symmetric METTS, U(1)-SYMETTS yields a reduction of CPU time by a factor between 4 and 10 for the parameters considered here (see Appendix \ref{app:cputime} for a more detailed assessment). Using a second-order Trotter decomposition, we set the time step $\tau_{\beta}=\tau_{\rm dyn}=0.05 J_2^{-1}$ and truncation error $s^{\rm tol}_{\beta} = 10^{-5}, s^{\rm tol}_{\rm dyn} = 5 \times 10^{-4}$  in the imaginary- and real-time evolution, respectively. We stop the real-time evolution at $t_{\rm max} = 50 J_2^{-1}$ and checked that calculations are not impaired by finite-size reflections on this time scale. This setup leads to a maximum bond dimension $D<600$ at $t_{\rm max}$ for all the time-evolved SYMETTS considered. To minimize finite-time effects, we here use a Gaussian broadening to perform the Fourier transform, leading to a frequency resolution $W \approx 0.16$ meV. As an alternative route, we had also tested linear prediction, but found that for this model its results were very sensitive to changes of the regularization parameter and the statistical window on the given time scale. Therefore while in principle, after significant further fine tuning, linear prediction may allow a systematic extrapolation to longer time scales to enhance spectral resolution, we did not further pursue this route.

Our results are displayed in \Fig{fig:DSSF_az}. Each column indicates a different magnetic field strength $H = 14, 20, 27$~T and each row corresponds to a  different  temperature $T = 0, 4.125, 8.25$~K.

 Let us first note that \Fig{fig:DSSF_az}(a) at zero temperature and $H=14$~T nicely reproduces all features of the DDMRG data used in \Ref{jeschke2011multistep} for a direct comparison with the experiment in \Ref{Rule_PRL_2008}. We observe a gapped system with a low-energy band dispersing along $k$, corresponding to the monomer excitations, and  a dimer branch at higher energies, whose dispersion is weakened by the competition of $J_1$ and $J_2$.\cite{jeschke2011multistep} The spectral weight in both branches is mainly distributed around $k=\pi$. Moreover, we find an additional excitation branch at $\omega > 3$~meV with only small spectral weight and almost no dispersion. In the experiment, this branch is shifted towards higher energies (by $\sim 1$ meV).\cite{Rule_PRL_2008}

Increasing the magnetic field has an effect on the position of both the monomer and dimer band (but not on their dispersion), which can be understood easily in an intuitive picture. As discussed in \Sec{sec:plateau_az},  the monomers are fully polarized in the entire plateau phase. Hence, exciting a monomer spin becomes increasingly expensive for larger magnetic fields, because a spin-flip is penalized by the additional Zeeman energy. Comparing the position of the monomer branch in \Figs{fig:DSSF_az}(a) and \ref{fig:DSSF_az}(c), the shift towards higher energies at $H=27$ T is fully captured by the change in the Zeeman term $g \mu_B \Delta H \approx 1.6$~meV. 

The magnetic field has the reversed effect on the dimer band, which is shifted to lower energies. Again the effect can be understood using the same line of arguments. Exciting a dimer singlet results in the break off of the singlet structure, allowing the dimer spins to polarize in the direction of $H$. At larger field strength, each excited dimer spin is therefore rewarded by a factor of  $(1/2) g \mu_B \Delta H$ from the Zeeman term. This fully accounts for the shift of the dimer branch to lower energies in \Figs{fig:DSSF_az}(a) and \ref{fig:DSSF_az}(c). At $H=20$ T, the system is approximately probed in the middle of the plateau phase (c.f.~\Fig{fig:azurite_plateau}). At this point, the band gap reaches a maximum, $\Delta E \approx 1$~meV, as the monomer branch has already moved to rather high energies while the dimer band is about to cross it, as illustrated in \Fig{fig:DSSF_az}(b). 

 \begin{figure*}
\centering
\includegraphics[width=\linewidth]{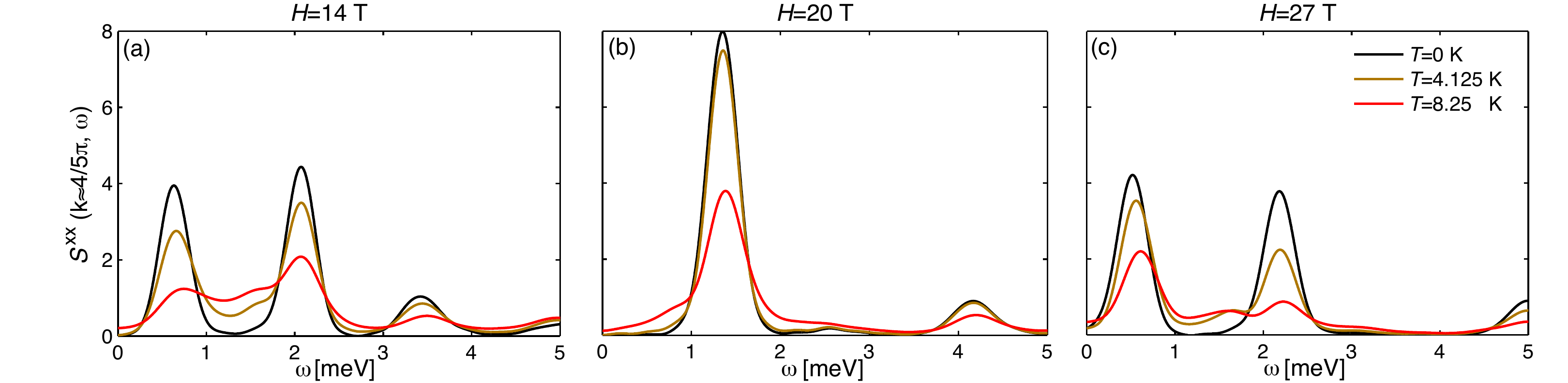}
\vspace{-15pt}
\caption{Transverse dynamic spin structure factor $S^{xx}(k,\omega)$  at $k\approx 4\pi/5$ of a generalized diamond chain model for azurite for three different field strengths in the plateau phase. At the edges of the plateau phase [(a) and (c)],  the large peaks indicating the monomer and dimer branches are already washed out at intermediate temperatures. Thermal fluctuations redistribute spectral weight in between the two excitation peaks in these cases. In the middle of the plateau phase (b) thermal broadening is almost not present at $T = 4.125$ K. Only at high temperatures is the height of the combined peak of monomer and dimer excitations significantly reduced.}
\label{fig:DSSF_Az_cross}
\end{figure*}

Based on this discussion, we can confirm the very distinct effects of temperature on the different points at the $\frac{1}{3}$ plateau and put the arguments given in \Sec{sec:plateau_az} on solid ground. For regions of the plateau where the magnetization decreases at finite temperature,  thermal fluctuations primarily  excite the monomers as this is energetically favorable. On the other hand, the thermal increase of the magnetization for larger magnetic fields observed in \Fig{fig:azurite_plateau}(b) can be understood in terms of the lowering of the dimer excitation energy due to the additional rewards in Zeemann energy, which has the opposite effect on the monomer band. 

\Fig{fig:DSSF_az}(d-i) displays the evolution of the spin excitations at finite temperature. The thermal broadening effects are strongly visible at $H=14$ and $27$ T, where the low energy bands are strongly smeared out even at intermediate temperatures. This is expected from the comparatively small band-gap at $T=0$ and \Fig{fig:azurite_plateau}(b), which shows strong effects of temperature on the magnetization in this regime. In contrast, thermal fluctuations have a much weaker effect at $H=20$~T, where the band gap is maximal. Indeed, comparing \Figs{fig:DSSF_az}(b) and \ref{fig:DSSF_az}(e), we  see almost no difference in  the distribution of the spectral weight. Only \Fig{fig:DSSF_az}(h) shows some thermal broadening, yet no new features arise in the spectrum. Again, this is in good agreement with the robustness of the magnetization for finite temperature in the middle of the plateau, as illustrated in \Fig{fig:DSSF_az}(b). These features become even more prominent when studying cross sections of \Fig{fig:DSSF_az}, i.e., the spin excitations for a specific momentum value. These are displayed in \Fig{fig:DSSF_Az_cross} for $k\approx4/5 \pi$. Again, we observe that  the large peaks indicating the monomer and dimer branches are already washed out at intermediate temperatures at the edges of the plateau phase [\Fig{fig:DSSF_Az_cross}(a) and \ref{fig:DSSF_Az_cross}(c)]. In both cases, thermal fluctuations strongly redistribute spectral weight in-between the two excitation peaks. In contrast, the height of the combined monomer and dimer excitation peaks  in the middle of the plateau phase is significantly reduced only at high temperatures [\Fig{fig:DSSF_Az_cross}(b)].

To conclude, our finite-temperature study of the spectra of the generalized diamond chain model for azurite fits in nicely with previous work\cite{jeschke2011multistep,Honecker_JCM_2011} and provides new insight in the plateau phase. We observe a crossing of the monomer and dimer branches with increasing magnetic field, which can very intuitively explain the effects of finite temperature on the plateau phase. Testing these features in neutron scattering experiments would provide additional information on the validity of the microscopic model for azurite. Such a study would be particularly enlightening in the context of the results provided by \Ref{rule2011dynamics}, which showed discrepancies of using an isotropic spin model to describe azurite in the regime of $H<H_{l,c}$, i.e., for fields below the plateau phase.

\FloatBarrier

\section{conclusion}
In this work, we have introduced an intuitive and easily implemented extension of the minimally entangled typical thermal state approach of \Ref{White_PRL_2011}, which allowed us to generate a METTS sample of symmetry eigenstates. We explicitly showed how to construct such a SYMETTS ensemble exploiting both the Abelian U$(1)_{\rm spin}$ and non-Abelian SU$(2)_{\rm spin}$ symmetry of spin-$\frac{1}{2}$ Heisenberg chains, without introducing strong autocorrelation effects in between the ensemble states. 

Whereas SYMETTS does not improve the numerical efficiency when calculating static observables as compared to METTS, the benefits of using symmetries come fully to the fore when calculating more complex dynamic quantities such as response functions. Here, most computational effort has to be put into the real-time evolution of each state in the sample and the gains of explicitly exploiting symmetries in the MPS simulations is enormous. We checked the validity of our approach for the dynamic spin structure factors of the XX and XXX Heisenberg chains and found that SYMETTS in principle is able to reach longer time scales than purification at low temperatures.

Moreover, we applied SYMETTS to study the finite-temperature excitation spectra of a generalized diamond chain model for the natural mineral azurite $\text{Cu}_3(\text{CO}_3)_2(\text{OH})_2$. Focusing on the plateau phase of the system, we found very distinct effects of temperature on the different points at the $\frac{1}{3}$ plateau, which are caused by the Zeeman term shifting the dimer and monomer branches in opposite directions. Our results fit in nicely with previous work\cite{jeschke2011multistep,Honecker_JCM_2011} and provide new insight in the plateau phase.

Interesting questions for future work involve the treatment of fermionic systems, where the symmetric ensemble states could be formulated in terms of a combination of ${\rm SU(2)}_{\rm charge}$ and ${\rm SU(2)}_{\rm spin}$ symmetries or their Abelian counterparts. For example, SYMETTS could be employed to study finite-temperature density profiles in interacting quantum-point contacts.\cite{Bauer_nature_2013} In this context, it would be particularly interesting to further explore the possibility of combining a real-time evolution to an MPO with local support in the Heisenberg picture, as briefly described at the end of \Sec{sec:dyn}. In principle, this would simplify combining METTS with the concept of time-translational invariance\cite{Barthel_NJP_2013} to double the maximum reachable time scale and could be a generally more efficient approach for finite-temperature response functions at low temperatures.

Finally we note as an outlook that the SYMETTS algorithm may also be entirely based within symmetry eigenstates, in that the non-symmetric sampling as described in this paper is fully replaced by Metropolis sampling. Based on the weights $P_{\bf q}$ above, necessarily, this must also include a proposal distribution to switch to neighboring symmetry sectors. To minimize rejection probability, this random walk towards neighboring symmetry sectors can be chosen temperature dependent. By definition, the Metropolis sampling also guarantees detailed balance. And, by rejecting certain higher-energy states, this may lead to reduced spread and hence enhanced convergence of computed observables. In this formulation, SYMETTS would also provide benefits for the calculation of static properties and might allow the finite-temperature treatment of 2D clusters.


\acknowledgments

We thank F.~Schwarz and M.~Stoudenmire  for insightful discussions.  This research was supported by
the Deutsche Forschungsgemeinschaft through the Excellence Cluster ``Nanosystems
Initiative Munich'', SFB/TR~12, SFB~631 and WE4819/1-1, WE4819/2-1 (AW). BB also acknowledges financial support from the BaCaTeC Grant 15 [2014-2].


\appendix
\section{Chebyshev expansion and METTS} \label{sec:CheMETTS}
Chebyshev expansion techniques have been successfully established as an alternative approach for the computation of spectral functions in the context of kernel polynomial methods.\cite{Weisse_RMP_2006} More recently, \Ref{Holzner_PRB_2011} introduced the Chebyshev expansion in the MPS formalism (CheMPS) to determine spectral properties at zero temperature. Based on this work, CheMPS has been applied to determine signatures of the Majorana fermion in the interacting Kitaev model,\cite{Thomale_PRB_2013} in the context of the interacting resonating level model,\cite{Schmitteckert_PRB_2014} and as impurity solver for single- and two-band DMFT calculations in combination with linear prediction.\cite{Ganahl_PRB_2014,AWolf_PRB_2014} In addition, CheMPS has been expanded towards finite-temperature calculations using a Liouvillian in a matrix-product purification framework.\cite{Tiegel_PRB_2014}

The question as to whether CheMPS is the most efficient method for computing spectral functions using MPS methods cannot be generally considered settled, as there is no one-to-one correspondence of CheMPS in its most efficient setup to real-time evolution. Nevertheless, the claim of \Ref{Holzner_PRB_2011} that CheMPS is significantly less expensive than tDMRG to obtain the same spectral information can no longer be supported.\cite{Wolf_PRB_2015}  We have not conducted a systematic comparison of both approaches, but in our experience CheMPS and tDMRG require similar computational effort when aiming for the same spectral resolution and employing an equal truncation criterion at zero temperature. CheMPS, though, offers a significant advantage over tDMRG as it allows better control over the broadening procedure of the spectral data.\cite{Holzner_PRB_2011,Schmitteckert_PRB_2014}

 In this context, it is worthwhile to explore the compatibility of SYMETTS and CheMPS. To this end, we start with the Fourier transform of the response function in \Eq{eq:response} 
\begin{eqnarray}
\mathcal{A}^{\hat{B}\hat{C}}_{\beta} (\omega)  &=&\int d\bar{\omega}  \langle   \delta(\bar{\omega}-\hat{H})   \hat{B}  \delta(\omega+\bar{\omega}-\hat{H}) \hat{C} \rangle_{\beta}\ . \quad\quad  \label{Eq:Metts_Che}
\end{eqnarray}
To compute the response function in this form, we  follow \Ref{Weisse_RMP_2006} and  expand both $\delta$-functions in terms of orthogonal Chebyshev polynomials of the first kind $T_m(\omega+\bar{\omega})$ and $T_n(\bar{\omega})$ before integrating over the frequency index $\bar{\omega}$  for every SYMETTS $|\phi_{\vq} \rangle$ in our sample. This ``double'' Chebyshev expansion involves Chebyshev moments of the type  $\mu^{\hat{B}\hat{C}}_{mn} = \langle T_m(\hat{H}')\hat{B}T_{n}(\hat{H'})\hat{C} \rangle_{\beta}$, where $\hat{H}'$ represents the Hamiltonian with a rescaled spectrum $\omega' \in [-1,1]$ to ensure the convergence of the Chebyshev recursion. This is usually achieved by using a linear rescaling with the parameters $a,b$
\begin{equation}
\hat{H}' = \frac{\hat{H}-b}{a}.
\end{equation}

\begin{figure}[t]
\centering
\includegraphics[width=\linewidth]{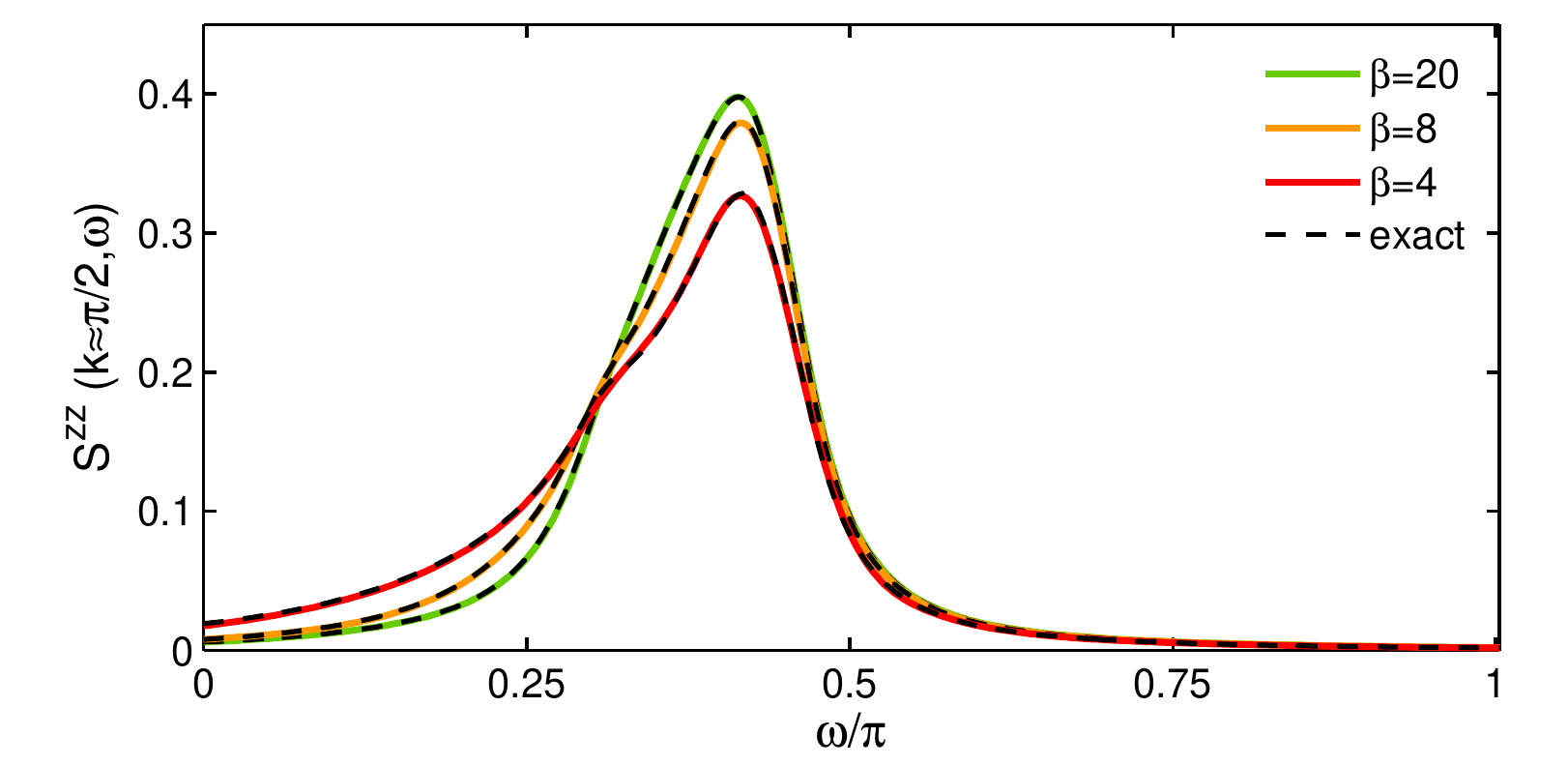}
\vspace{-10pt}
\caption{CheMETTS calculation for spin structure factor of the XX model with $N=50$, $s_{\rm tol}^{\rm dyn} = 5 \times 10^{-4}$, $N_{\rm Che} = 300$, $\eta=0.1$ and $M=300$. For all considered inverse temperatures, we find excellent agreement with the exact result. However, note that the required numerical resources clearly exceed those used in \Fig{fig:HB_XX_DSSF_tDMRG}, where tDMRG was employed for a system with twice as many spins! }
\label{fig:HB_XX_DSSF_CheMETTS}
\end{figure}

The  moments $\mu_{mn}$ are determined by calculating a first set of Chebyshev vectors up to the desired expansion order $N_{\rm Che}$ via the recursion relation
\begin{eqnarray}
|t_m\rangle &=& 2\hat{H}' |t_{m-1}\rangle - |t_{m-2}\rangle, \nonumber \\
 |t_0\rangle &=&  |\phi_{\vq} \rangle, \ |t_1\rangle = \hat{H}' |t_0\rangle, 
\end{eqnarray}
and keeping it in storage. Then, we iteratively obtain a second set of vectors $|\tilde{t}_n\rangle$ using a different starting vector $|\tilde{t}_0\rangle = \hat{C} |\phi_{\vq} \rangle$. For each $|\tilde{t}_n\rangle$, we compute the overlap $\mu_{mn} = \langle t_m | \hat{B}| \tilde{t}_n \rangle$ for $m=0,1,...,N_{\rm Che}-1$. 

The sample average of the Chebyshev moments is then used to compute the finite-temperature response function in frequency space.

To this end, we work with a finite broadening $\eta$ instead of the usual kernel approach for $\delta(\omega' + \bar{\omega}' -\hat{H}')$ to remove the artificial ``Gibbs'' oscillations  caused by finite expansion order from the spectral data.\cite{Schmitteckert_PRB_2014} Note that the broadening has to be performed for the $\delta$-function containing the external frequency index only as $\bar{\omega}'$ is integrated out to obtain the final result. The finite-temperature response function in \Eq{Eq:Metts_Che} then takes the form
\begin{eqnarray}
\mathcal{A}^{\hat{B}\hat{C}}_{\beta} (\omega) 
&=& \frac{1}{a} \sum_{m,n=0}^{N_{\rm Che}-1}  \mu_{mn} (2-\delta_{m0}) \nonumber \\
&&  \int {\rm d}\bar{\omega}'  \frac{1}{\pi \sqrt{1-\bar{\omega}'^2}}  T_m(\bar{\omega}') \alpha_n(z),\label{eq:directA}
\end{eqnarray} 
with $z = (\omega' + \bar{\omega}') + i \eta /a$ and  $\alpha_n$ given by\cite{Schmitteckert_PRB_2014}
\begin{equation}
\alpha_n (z) = \frac{2/(1+\delta_{n0})}{(z)^{n+1}(1+\sqrt{z^2}\sqrt{z^2 -1}/z^2)^n\sqrt{1-1/z^2}}.
\end{equation}
In principle, this approach represents an alternative to the combination of tDMRG plus Fourier transform, which we have applied in the main part of this work. We illustrate this in \Fig{fig:HB_XX_DSSF_CheMETTS}, where we used U(1)-SYMETTS and a double Chebyshev expansion to compute the dynamic spin structure factor of the XX model showing excellent agreement with exact calculations (dashed lines). However, the Chebyshev approach in the METTS formalism involves significantly higher computational costs than the real-time evolution, since, in contrast to $T=0$ CheMPS, (i) the full set of Chebyhsev vectors $|t_m\rangle$ has to be stored throughout the entire calculation, and (ii) the number of moments increases from $N_{\rm Che}$ to $N_{\rm Che}^2$, also squaring the number of MPS overlaps to be calculated.

Therefore, we conclude that the combination of CheMPS and METTS is not a competitive alternative to real-time evolution for calculating spectra at finite temperature, since the advantage of the more controlled broadening procedure does not outweigh the drastically enhanced numerical costs involved in the double Chebyshev expansion.\\

\section{Numerical efficiency of SYMETTS for azurite}
\label{app:cputime}
\begin{figure}[b]
\centering
\includegraphics[width=\linewidth]{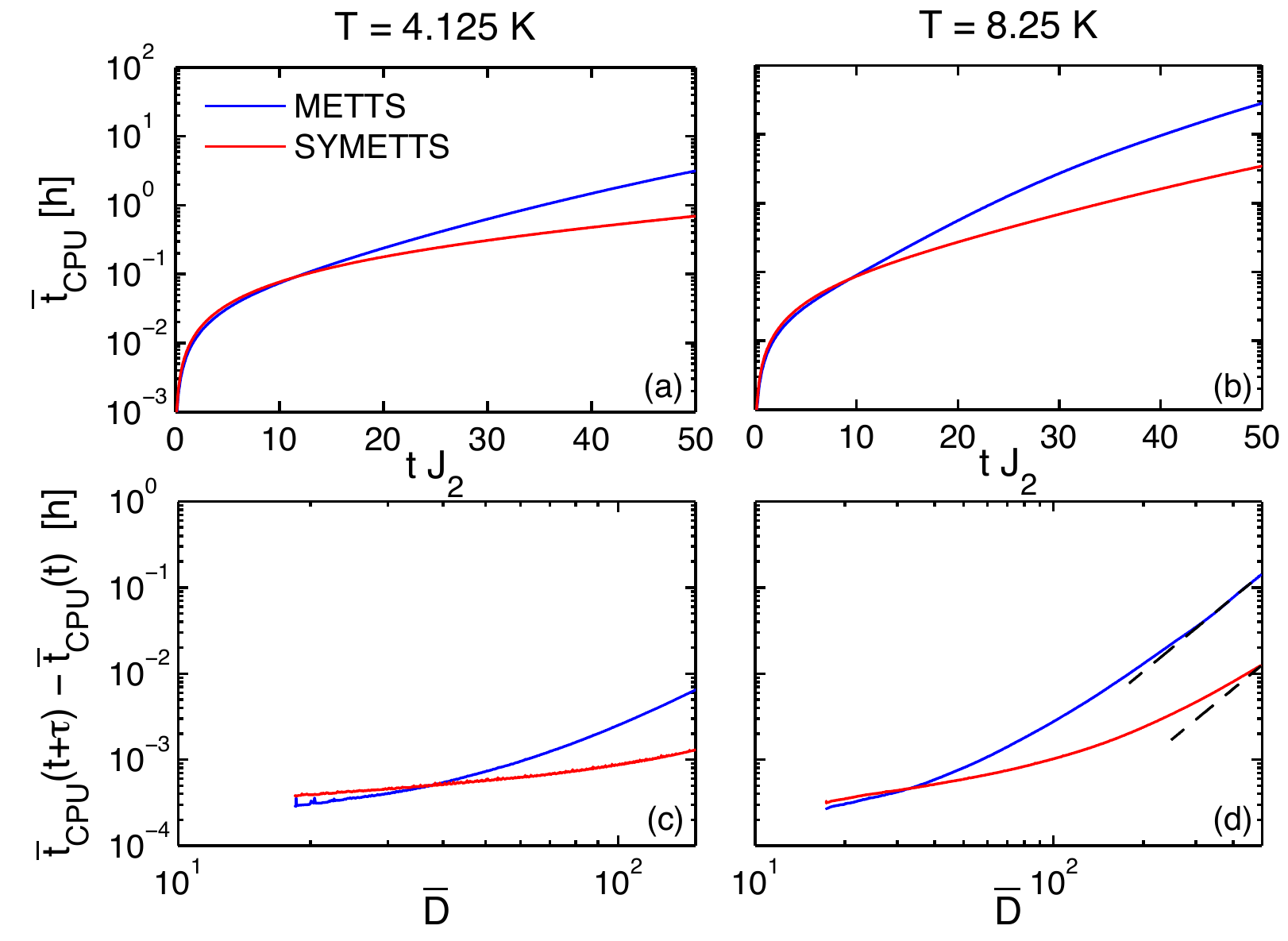}
\vspace{-20pt}
\caption{(a,b) display the numerical performance of SYMETTS and METTS in terms of the average cumulative CPU time $\bar{t}_{\rm CPU}$ as a function of $tJ_2$, when carrying out the real-time evolution to determine the dynamic spin structure factor $S^{xx}(k,\omega)$ of azurite in \Eq{eq:DSSF_az} using the parameters $N=90$, $H=14$ T, and $k=4/5 \pi$. (c,d) show the average computation time of a \emph{single} time step $\tau$  as a function of $\bar{D}$. We note that the ensemble states have a $\bar{D}\approx20$ at both temperatures after the initial imaginary-time evolution. The dashed lines in \Fig{fig:tCPU}(d) are guide to the eye illustrating the $\bar{t}_{\rm CPU} \sim D^3$ scaling of the CPU time for larger bond dimensions. The employed tDMRG setup is  described in  \Sec{sec:Az_DSSF}.}
\label{fig:tCPU}
\end{figure}

Here we assess the numerical performance of SYMETTS and the existing METTS approach on an explicit example. We focus on the average cumulative CPU time $\bar{t}_{\rm CPU}$ required to carry out  the real-time evolution of one ensemble state up to $t \le t_{\rm max} = 50 J_2^{-1}$ when determining the dynamic spin structure factor $S^{xx}(k,\omega)$ of azurite in \Eq{eq:DSSF_az}. For simplicity, we choose the same model parameters as in \Fig{fig:DSSF_Az_cross}(a), namely $N=90$, $H=14$ T, and $k=4/5 \pi$, and in \Figs{fig:tCPU}(a,b) display the resulting average cumulative CPU times for the two temperatures $T=4.125, 8.25$ K. Using the tDMRG setup described in \Sec{sec:Az_DSSF}, each calculation was performed  on a single core Xeon E5-2670v2 (2.50 GHz) machine with 4GB memory.

The explicit implementation of the U(1) symmetry in the SYMETTS ensemble states clearly enhances the numerical efficiency, resulting in an average reduction of CPU time by a factor of 4 for $T=4.125$ K in comparison to a non-symmetric METTS sample [c.f.~\Fig{fig:tCPU}(a)]. The efficiency gain increases to a factor of almost 10 for  $T=8.25$ K in \Fig{fig:tCPU}(b), since states with a larger bond dimension profit even more from the exploitation of the spin symmetry. This is illustrated in \Fig{fig:tCPU}(c,d), where we show the average computation time of a \emph{single} time step $\tau$  as a function of the average maximum bond dimension $\bar{D}$.  The additional imaginary-time evolution necessary for the generation of the SYMETTS requires on average only 25 and 12 seconds for $T=4.125, 8.25$ K, respectively.  Thus, the overhead costs of the generating the SYMETTS sample are clearly negligible compared to the total computation time.

In addition to benefits in terms of memory requirement, SYMETTS  enables us to reduce the CPU time necessary to compute the dynamic spin structure factor for various momenta and magnetic fields presented in \Fig{fig:DSSF_az} from $\mathcal{O}(10^{6})$ to $\mathcal{O}(10^{5})$ hours. Consequently, when running $400$ CPUs in parallel, SYMETTS generates this data in roughly one week, whereas the same calculation would require almost three months in the original METTS formulation. Note that the factor of 10 gained in numerical efficiency by implementing the U(1) spin symmetry in simple spin-chain models has also been reported in \Ref{Singh_PRB_2012} and [\onlinecite{Kennes_2014}] in the context of iTEBD and tDMRG, respectively. Even larger benefits can be achieved when studying models with multiple Abelian or non-Abelian symmetries.

%

\end{document}